\documentclass[a4,11pt]{article}
\usepackage{graphicx} 
\usepackage{float}
\usepackage{amsmath,amsthm}
\usepackage{amsfonts}
\usepackage{latexsym}
\usepackage{amssymb}
\usepackage{setspace}
\usepackage{comment}

\makeatletter
 
  \@addtoreset{equation}{section}
 \makeatother

\begin{document}
\title{Asymptotic Expansion for Forward-Backward SDEs\\ with Jumps~\footnote{
Forthcoming in {\it Stochastics}. 
All the contents expressed in this research are solely those of the authors and do not represent any views or 
opinions of any institutions. The authors are not responsible or liable in any manner for any losses and/or damages caused by the use of any contents in this research.
}
}

\author{Masaaki Fujii\footnote{Quantitative Finance Course, Graduate School of Economics, The University of Tokyo. }
~~\&~~Akihiko Takahashi\footnote{Quantitative Finance Course, Graduate School of Economics, The University of Tokyo.} 
}
\date{ \small 
This version: 7 September, 2018}
\maketitle



\newtheorem{definition}{Definition}[section]
\newtheorem{assumption}{Assumption}[section]
\newtheorem{condition}{$[$ C}
\newtheorem{lemma}{Lemma}[section]
\newtheorem{proposition}{Proposition}[section]
\newtheorem{theorem}{Theorem}[section]
\newtheorem{remark}{Remark}[section]
\newtheorem{example}{Example}[section]
\newtheorem{corollary}{Corollary}[section]
\def\n{{\bf n}}
\def\A{{\bf A}}
\def\B{{\bf B}}
\def\C{{\bf C}}
\def\D{{\bf D}}
\def\E{{\bf E}}
\def\F{{\bf F}}
\def\G{{\bf G}}
\def\H{{\bf H}}
\def\I{{\bf I}}
\def\J{{\bf J}}
\def\K{{\bf K}}
\def\L{{\bf L}}
\def\M{{\bf M}}
\def\N{{\bf N}}
\def\O{{\bf O}}
\def\P{{\bf P}}
\def\Q{{\bf Q}}
\def\R{{\bf R}}
\def\S{{\bf S}}
\def\T{{\bf T}}
\def\U{{\bf U}}
\def\V{{\bf V}}
\def\W{{\bf W}}
\def\X{{\bf X}}
\def\Y{{\bf Y}}
\def\Z{{\bf Z}}
\def\cala{{\cal A}}
\def\calb{{\cal B}}
\def\calc{{\cal C}}
\def\cald{{\cal D}}
\def\cale{{\cal E}}
\def\calf{{\cal F}}
\def\calg{{\cal G}}
\def\calh{{\cal H}}
\def\cali{{\cal I}}
\def\calj{{\cal J}}
\def\calk{{\cal K}}
\def\call{{\cal L}}
\def\calm{{\cal M}}
\def\caln{{\cal N}}
\def\calo{{\cal O}}
\def\calp{{\cal P}}
\def\calq{{\cal Q}}
\def\calr{{\cal R}}
\def\cals{{\cal S}}
\def\calt{{\cal T}}
\def\calu{{\cal U}}
\def\calv{{\cal V}}
\def\calw{{\cal W}}
\def\calx{{\cal X}}
\def\caly{{\cal Y}}
\def\calz{{\cal Z}}
%
\def\sskip{\hspace{0.5cm}}
\def\simleq{ \raisebox{-.7ex}{\em $\stackrel{{\textstyle <}}{\sim}$} }
\def\leqsim{ \raisebox{-.7ex}{\em $\stackrel{{\textstyle <}}{\sim}$} }
\def\ep{\epsilon}
\def\half{\frac{1}{2}}
\def\iku{\rightarrow}
\def\Iku{\Rightarrow}
\def\ikup{\rightarrow^{p}}
\def\inclusion{\hookrightarrow}
\def\cadlag{c\`adl\`ag\ }
\def\up{\uparrow}
\def\down{\downarrow}
\def\doti{\Leftrightarrow}
\def\douti{\Leftrightarrow}
\def\dochi{\Leftrightarrow}
\def\douchi{\Leftrightarrow}%
\def\yy{\\ && \nonumber \\}
\def\y{\vspace*{3mm}\\}
\def\nn{\nonumber}
\def\be{\begin{equation}}
\def\ee{\end{equation}}
\def\bea{\begin{eqnarray}}
\def\eea{\end{eqnarray}}
\def\beas{\begin{eqnarray*}}
\def\eeas{\end{eqnarray*}}
%
\def\hd{\hat{D}}
\def\hv{\hat{V}}
\def\hsd{{\hat{d}}}
\def\hx{\hat{X}}
\def\hsx{\hat{x}}
\def\bsx{\bar{x}}
\def\bsd{{\bar{d}}}
\def\bx{\bar{X}}
\def\ba{\bar{A}}
\def\bb{\bar{B}}
\def\bc{\bar{C}}
\def\bv{\bar{V}}
\def\balpha{\bar{\alpha}}
\def\bbalpha{\bar{\bar{\alpha}}}
\def\combi{\l(\begin{array}{c}\alpha\\ \beta \end{array}\r)}
\def\f{^{(1)}}
\def\s{^{(2)}}
\def\ss{^{(2)*}}
\def\l{\left}
\def\r{\right}
\def\a{\alpha}
\def\b{\beta}
\def\L{\Lambda}


\def\E{{\bf E}}
\def\P{{\bf P}}
\def\Q{{\bf Q}}
\def\R{{\bf R}}

\def\cadlag{{c\`adl\`ag~}}

\def\calf{{\cal F}}
\def\calp{{\cal P}}
\def\calq{{\cal Q}}
\def\wtW{\widetilde{W}}
\def\wtB{\widetilde{B}}
\def\wtPsi{\widetilde{\Psi}}
\def\wt{\widetilde}
\def\mbb{\mathbb}
\def\opi{{\pi^*}}
\def\odel{{\del^*}}
\def\pii{{\pi^i}}
\def\pidel{{\pi,\del}}
\newcommand{\bvec}[1]{\mbox{\boldmath $#1$}}
\def\bpi{\bvec{\pi}}
\def\bX{\bvec{X}}
\def\bx{\bvec{x}}
\def\by{\bvec{y}}
\def\bee{\bvec{e}}
\def\bbs{\bvec{b}}
\def\bS{\bvec{S}}
\def\bdel{\bvec{\del}}
\def\bPhi{\bvec{\Phi}}
\def\bPsi{\bvec{\Psi}}
\def\bbeta{\bvec{\beta}}
\def\bl{\bvec{l}}
\def\bdel{\bvec{\del}}
\def\bTheta{\bvec{\Theta}}
\def\btheta{\bvec{\theta}}

\def\LDis{\frac{\bigl.}{\bigr.}}
\def\ep{\epsilon}
\def\del{\delta}
\def\Del{\Delta}
\def\part{\partial}
\def\wh{\widehat}
\def\bsigma{\bar{\sigma}}
\def\yy{\\ && \nonumber \\}
\def\y{\vspace*{3mm}\\}
\def\nn{\nonumber}
\def\be{\begin{equation}}
\def\ee{\end{equation}}
\def\bea{\begin{eqnarray}}
\def\eea{\end{eqnarray}}
\def\beas{\begin{eqnarray*}}
\def\eeas{\end{eqnarray*}}
\def\l{\left}
\def\r{\right}

\def\bull{$\bullet~$}

\newcommand{\Slash}[1]{{\ooalign{\hfil/\hfil\crcr$#1$}}}
\vspace{0mm}

\begin{abstract}
This work provides a  semi-analytic approximation method for decoupled forward-backward SDEs (FBSDEs) with  jumps. 
In particular, we construct an asymptotic expansion method for FBSDEs driven by the
random Poisson measures with $\sigma$-finite compensators as well as the standard Brownian motions
around the small-variance limit of the forward SDE.
We provide a 
semi-analytic solution technique as well as its error estimate
for which we only need to solve essentially a system of linear ODEs.
In the case of a finite jump measure with a bounded intensity, the method can also handle state-dependent and hence non-Poissonian
jumps, which are quite relevant for many practical applications.


\end{abstract}
\vspace{2mm}

\section{Introduction}
Since it was introduced by Bismut (1973)~\cite{Bismut} and
Pardoux \& Peng (1990)~\cite{Pardoux-Peng}, the backward stochastic differential equation (BSDE)
has attracted many mathematicians because of its deep connections to 
non-linear partial differential equations.
There now exist excellent reviews such as El Karoui \& Mazliak (eds.) (1997)~\cite{ElKaroui-Mazliak},
Ma \& Yong (2000)~\cite{Ma-Yong}, and 
Pardoux \& Rascanu (2014)~\cite{Pardoux-Rascanu} for interested readers.
BSDEs also have a wide variety of applications to financial as well as operational problems;
El Karoui et al. (1997)~\cite{ElKaroui}, Lim (2004)~\cite{Lim},
Jeanblanc \& Hamad\`ene (2007)~\cite{Jeanblanc},
Cvitani\'c \& Zhang (2013)~\cite{Cvitanic}, 
Touzi (2013)~\cite{Touzi} and Cr\'epey, Bielecki \& Brigo (2014)~\cite{Crepey} to mention only a few.
As for BSDEs with jumps, see for example, Barles, Buckdahn \& Pardoux (1997)~\cite{Barles},
Royer (2006)~\cite{Royer}, Crepey \& Matoussi (2008)~\cite{Crepey-Matoussi},
Morlais (2010)~\cite{Morlais}, Delong (2013)~\cite{Delong} and Quenez \& Sulem (2013)~\cite{Quenez-S}.

The last financial crisis and a bunch of new 
regulations that followed have made various problems involving BSDEs such as XVAs, risk measures,  optimal executions in illiquid markets
and the development for their efficient numerical computation scheme the central issues in the financial industry.
Although the backward Monte-Carlo simulation scheme has been proposed and studied by many researchers such as, Bouchard \& Touzi (2004)~\cite{Bouchard-Touzi}, Zhang (2004)~\cite{ZhangJ}, Gobet et al. (2004)~\cite{Gobet}
and Bender \& Denk (2007)~\cite{Bender-Denk} for continuous BSDEs,
and Bouchard \& Elie (2008)~\cite{Bouchard-Elie} for BSDEs with jumps,  it has not yet become a standard tool for practitioners
due to its computational burden.
In particular, we can only find  simple one-dimensional examples using the Poisson process instead of a 
random measure in the existing literature. See, for example, Elie (2006)~\cite{Elie} and Lejay et.al. (2014)~\cite{Lejay}.
See also the discussion in \cite{Crepey} and  
Cr\'epey \& Song~(2015)~\cite{Crepey-Song} regarding the problems in the existing computation scheme when
applied to practical problems~\footnote{In \cite{Crepey-Song},
the authors successfully applied the asymptotic expansion method 
proposed in \cite{FT-analytic, FT-particle} to a collateralized debt obligation with 120 underlying names
to evaluate credit/funding valuation adjustments.}.
Moreover, in certain applications such as mean-variance hedging and multiple dependent defaults, 
the solution of one BSDE appears in the driver of another BSDE~\footnote{See
Mania \& Tevzadze (2003)~\cite{Mania}, Pham (2010)~\cite{Pham} and Fujii (2015)~\cite{Fujii-mm}
for concrete examples.}. In such a case,  deriving an analytic approximation for the first BSDE 
seems the only possibility to obtain a numerical result within reasonable computational time.

As one possible approach to these problems, the current work contributes by providing a straightforward semi-analytic
approximation method for BSDEs with jumps, which are especially difficult and time-consuming to evaluate 
by the standard Monte-Carlo scheme.
We develop an asymptotic expansion method for decoupled forward-backward SDEs (FBSDEs)
with Lipschitz drivers and the Poisson random measures in addition to the standard Brownian motions.
We propose an expansion around a small-variance limit of the forward SDE.
It starts from solving a non-linear ODE corresponding to the BSDE where every forward component 
is replaced by the deterministic mean process.
Every higher order approximation yields a linear FBSDE,
which can be solved semi-analytically essentially by a system of linear ODEs.
More precisely, the approximate solution of the BSDE including the martingale components
is explicitly given by a polynomial in the stochastic flows of the  forward process
whose coefficients can be computed by the linear ODEs.

In order to justify the approximation method  and its error estimate,
we use the results of Kruse \& Popier (2015)~\cite{Kruse} for a priori estimates and the existence of unique 
$\mbb{L}^p$-solution of BSDEs with jumps,
Delong \& Imkeller (2010)~\cite{Delong-Imkeller}
and Delong~\cite{Delong} for the representation theorem based on the Malliavin's derivative, 
as well as the idea of Pardoux \& Peng (1992)~\cite{Pardoux-Peng-lec} and Ma \& Zhang (2002)~\cite{Ma-Zhang}
for controlling the sup-norm of the martingale integrands of the BSDEs.
In the case of a finite jump measure with a bounded intensity, the method can also be applied to
a system with state-dependent and hence non-Poissonian jumps, which are quite relevant for many practical applications.
A closed-form  expression of the approximation up to an arbitrary higher order term is available when
 the forward SDE belongs to (time-inhomogeneous) exponential L\'evy type.
The current work also serves as a justification of a polynomial 
expansion method proposed in Fujii (2015)~\cite{Fujii-poly} for a certain class of models,
which provides a couple of interesting numerical examples.
\\

The organization of the paper is as follows:
Section~2 gives some preliminaries, Section~3 the setup of the interested FBSDEs and 
the representation theorem based on Malliavin's derivative, Section 4 the asymptotic expansion and its error estimate,
and finally Section 5 gives the concrete implementation of the scheme.
Appendices~A and B summarize the relevant a priori estimates,
and Appendix~C provides the smooth approximation theorem for the FBSDEs, which justifies the assumptions used in 
the main text.

\begin{remark}
As for forward SDEs, the asymptotic expansion method around a small-variance limit has been
applied to a variety of financial problems. It has been shown, in various numerical examples,
that the first few terms of expansion are enough to achieve accurate approximation
for option pricing  with typical volatilities ranging from 10\% to 20\% 
and maturities up to a few years.
See a review Takahashi (2015)~\cite{T-review} for the details 
and a comprehensive list of literature.  
\end{remark}

\begin{remark}
The current work can be extended in couple of ways.
Firstly, based on the result of Fujii \& Takahashi (2017)~\cite{FT-qg-jumps},
a similar asymptotic expansion may be justified for a BSDE with a quadratic-exponential growth driver
and a bounded terminal condition.  This would be done by replacing the estimates of the standard Lipschitz BSDEs with those
of local Lipschitz BSDEs with $\mbb{H}^2$-BMO coefficients.
It may also be possible to develop a sub-stepping scheme similar to those in Fujii (2014)~\cite{F-Filtering} and 
Takahashi \& Yamada (2015)~\cite{TY-AAP}, which can handle higher volatilities and longer maturities.
See an initial attempt in a diffusion setup with quadratic growth driver by Fujii \& Takahashi (2016)~\cite{FT-short-term}.
\end{remark}

\section{Preliminaries}
\label{sec-Preliminaries}
\subsection{General Setting}
$T>0$ is some bounded time horizon.
The space $(\Omega_W,\calf_W,\mbb{P}_W)$
is the usual canonical space for an $l$-dimensional Brownian motion equipped with the Wiener measure $\mbb{P}_W$.
$(\Omega_\mu,\calf_\mu,\mbb{P}_\mu)$ denotes a product of 
canonical spaces $\Omega_\mu:=\Omega_\mu^1\times \cdots\times \Omega_\mu^k$,
$\calf_\mu:=\calf_\mu^1\times \cdots \times \calf_\mu^k$ and $\mbb{P}_\mu^1\times \cdots \times \mbb{P}_\mu^k$ with some integer $k\geq 1$,
on which each $\mu^i$ is a Poisson measure with a compensator $\nu^i(dz)dt$. Here, $\nu^i(dz)$ is a $\sigma$-finite 
measure on $\mbb{R}_0:=\mbb{R}\backslash\{0\}$ satisfying $\int_{\mbb{R}_0} |z|^2 \nu^i(dz)<\infty$.
Throughout the paper, we work on the filtered probability space 
$(\Omega, \calf, \mbb{F}, \mbb{P})$, where the  space $(\Omega,\calf,\mbb{P})$
is the product of the canonical spaces $(\Omega_W\times  \Omega_\mu, \calf_W\times\calf_\mu,\mbb{P}_W\times \mbb{P}_\mu)$,
and that the filtration $\mbb{F}:=(\calf_t)_{t\in[0,T]}$ is the 
canonical filtration completed for $\mbb{P}$ and satisfying the usual conditions.
In this construction, $(W,\mu^1,\cdots,\mu^k)$ are independent 
and it is well-know that {\it the predictable representation property} holds~\footnote{See, for example,  Chapter XIII in \cite{He-Wang-Yan}. If one assumes the predictable representation property, this construction is irrelevant.}. 
We use a vector notation $\mu(\omega, dt,dz):=(\mu^1(\omega, dt,dz^1), \cdots, \mu^k(\omega, dt,dz^k))$.
The compensated Poisson measure is denoted by $\wt{\mu}:=\mu-\nu$.
We represent the $\mbb{F}$-predictable $\sigma$-field on $\Omega\times [0,T]$ by $\calp$.

\subsection{Notation}
Let $C_p$ denote a generic constant, which may change line by line, 
depending on $p$, $T$ and the Lipschitz constants and the bounds of the relevant functions.
For any integer $r\geq 1$, let us introduce a sup-norm for a $\mbb{R}^r$-valued function $x:[0,T]\rightarrow \mbb{R}^r$ as
\be
||x||_{[a,b]}:=\sup\{|x_t|,t\in[a,b]\}~ \nn
\ee
and write $||x||_t:=||x||_{[0,t]}$.
\\\\
Let us introduce the following spaces for stochastic processes for $p\geq 2$:\\
\bull $\mbb{S}^p_r[s,t]$ is the set of $\mbb{R}^r$-valued adapted \cadlag
processes $X$ such that
\be
||X||_{\mbb{S}^p_r[s,t]}:=\mbb{E}\left[||X(\omega)||_{[s,t]}^p\right]^{1/p}<\infty~. \nn
\ee
\bull $\mbb{H}^p_r[s,t]$ is the set of progressively measurable $\mbb{R}^r$-valued processes $Z$ 
such that
\be
||Z||_{\mbb{H}^p_r[s,t]}:=\mbb{E}\left[ \Bigl(\int_s^t |Z_u|^2 du\Bigr)^{p/2}\right]^{1/p}<\infty. \nn 
\ee
\bull $\mbb{H}^p_{r,\nu}[s,t]$ is the set of functions $\psi=\{(\psi)_{i,j},1\leq i\leq r, 1\leq j\leq k\}$,
$(\psi)_{i,j}: \Omega\times [0,T]\times \mbb{R}_0 \rightarrow \mbb{R}$
which are $\calp\times \calb(\mbb{R}_0)$-measurable and satisfy
\bea
||\psi||_{\mbb{H}^p_{r,\nu}[s,t]}:=\mbb{E}\left[\Bigl(\sum_{i=1}^k\int_s^t \int_{\mbb{R}_0}|\psi_u^{\cdot,i}(z)|^2\nu^i(dz)du
\Bigr)^{p/2}\right]^{1/p}<\infty. \nn
\eea
For simplicity, we use the notation $(E,\cale):=(\mbb{R}_0^k,\calb(\mbb{R}_0)^k)$ and denote 
the above maps $\{(\psi)_{i,j},1\leq i\leq r,1\leq j\leq k\}$
by $\psi:\Omega\times [0,T]\times E\rightarrow \mbb{R}^{r\times k}$ and say $\psi$ is $\calp\times \cale$-measurable 
without referring to each component.
We also use the notation such that
\bea
\int_s^t\int_E \psi_u(z)\wt{\mu}(du,dz):=\sum_{i=1}^k \int_s^t \int_{\mbb{R}_0}\psi_u^i(z)\wt{\mu}^i(du,dz)~ \nn
\eea
for simplicity. The similar abbreviation is used also for the integral with $\mu$ and $\nu$.
When we use $E$ and $\cale$, one should always interpret it in this way so that the integral
with the $k$-dimensional Poisson measure does make sense. On the other hand, when we use the range $\mbb{R}_0$
with the integrators $(\wt{\mu},\mu, \nu)$, for example, 
\bea
\int_{\mbb{R}_0}\psi_u(z)\nu(dz):=\Bigl(\int_{\mbb{R}_0}\psi_u^i(z)\nu^i(dz)\Bigr)_{1\leq i\leq k} \nn
\eea
we interpret it as a $k$-dimensional vector.\\
\bull $\calk^p[s,t]$ is the set of functions $(Y,Z,\psi)$ in the space $\mbb{S}^p[s,t]\times \mbb{H}^p[s,t]\times \mbb{H}^p_\nu[s,t]$
with the norm defined by 
\be
||(Y,Z,\psi)||_{\calk^p[s,t]}:=\Bigl(||Y||_{\mbb{S}^p[s,t]}^p+||Z||_{\mbb{H}^p[s,t]}^p+||\psi||_{\mbb{H}^p_\nu[s,t]}^p\Bigr)^{1/p}\nn .
\ee
\bull $\mbb{L}^2(E,\cale,\nu:\mbb{R}^r)$
is the set of $\mbb{R}^{r\times k}$-valued $\cale$-measurable functions $U$ satisfying
\bea
||U||_{\mbb{L}^2(E)}&:=&\Bigl(\int_E |U(z)|^2\nu(dz)\Bigr)^{1/2}\nn \\
&:=& \Bigl(\sum_{i=1}^k \int_{\mbb{R}_0} |U^{\cdot,i}(z)|^2\nu^i(dz)\Bigr)^{1/2}<\infty~. \nn
\eea
We frequently omit the subscripts for its dimension $r$ and the time interval $[s,t]$ when they are
obvious in the context. \\

We use the notation of partial derivatives such that
\bea
&&\part_\ep:=\frac{\part}{\part\ep}, \quad \part_x:=(\part_{x_1},\cdots, \part_{x_d}):=\Bigl(\frac{\part}{\part x_1},\cdots,\frac{\part}{\part x_d}\Bigr) \nn \\
&&\part_x^2:=\part_{x,x}:=\Bigl(\frac{\part^2}{\part x_i \part x_j}\Bigr)_{i,j=\{1,\cdots,d\}} \nn
\eea
and similarly for every higher order derivative without detailed indexing. We suppress the obvious summation of indexes 
throughout the paper for notational simplicity.

\section{Forward and Backward SDEs}
\subsection{The setup and some standard results}
We work in the filtered probability space $(\Omega,\calf,\mbb{F},\mbb{P})$ defined in the last section.
Let us introduce a $d$-dimensional forward SDE of $(X_s^{t,x,\ep}, s\in[t,T])$
with the initial data $(t,x)\in[0,T]\times \mbb{R}^d$ and a small perturbation parameter $\ep\in[0,1]$,
and an $m$-dimensional Markovian BSDE driven by $X^{t,x,\ep}$:
\bea
&&X_s^{t,x,\ep}=x+\int_t^s b(r,X_r^{t,x,\ep},\ep)dr+\int_t^s \sigma(r,X_r^{t,x,\ep},\ep)dW_r+\int_t^s \int_E \gamma(r,X_{r-}^{t,x,\ep},z,\ep)\wt{\mu}(dr,dz)\nn \\
\label{eq-X}\\
&&Y_s^{t,x,\ep}=\xi(X_T^{t,x,\ep})+
\int_s^T f\Bigl(r,X_r^{t,x,\ep},Y_r^{t,x,\ep},Z_r^{t,x,\ep},\int_{\mbb{R}_0}\rho(z)\psi_r^{t,x,\ep}(z)\nu(dz)\Bigr)dr\nn \\
&&\quad-\int_s^T Z_r^{t,x,\ep}dW_r-\int_s^T \int_E \psi_r^{t,x,\ep}(z)\wt{\mu}(dr,dz),
\label{eq-Y}
\eea
for $s\in[t,T]$. Here, $b: [0,T]\times \mbb{R}^d\times \mbb{R}\rightarrow \mbb{R}^d$,
$\sigma:[0,T]\times \mbb{R}^d\times \mbb{R}\rightarrow \mbb{R}^{d\times l}$ and
$\gamma:[0,T]\times \mbb{R}^d\times E\times \mbb{R}\rightarrow \mbb{R}^{d\times k}$ for the forward SDE
and
$\xi:\mbb{R}^d\rightarrow \mbb{R}^m$, $f:[0,T]\times \mbb{R}^d\times \mbb{R}^m\times
\mbb{R}^{m\times l}\times \mbb{R}^{m\times k}\rightarrow \mbb{R}^m$ and
$\rho:E\rightarrow \mbb{R}^k$ for the BSDE are measurable functions.

We shall specify the dependence of $(b,\sigma, \gamma)$ in $\ep$ more explicitly in Section~\ref{sec-implementation},
where we arrange it so that $X^{t,x,\ep}$ becomes deterministic process in the limit of $\ep\rightarrow 0$.
The main goal of the current paper is to obtain the Taylor expansion of the solution $(X^{t,x,\ep},Y^{t,x,\ep},Z^{t,x,\ep},
\psi^{t,x,\ep})$ around $\ep=0$ and the associated error estimates.
Let us fix the order of the highest expansion by $n_{\rm max} ~(\in \mbb{N})$ in the reminder of the paper.
For notational simplicity, let us define $n^{\rm ae}:=n_{\rm max}+2$.~\footnote{The additional factor $+2$ (instead of $+1$) arises 
basically from  the need to bound the approximation error for the control variables $(Z,\psi)$.}
\\

Let us also introduce the function $\eta:\mbb{R}\rightarrow \mbb{R}$ by $\eta(z):=1\wedge |z|$.
Now, we make the following assumptions.  
\begin{assumption}
\label{assumption-X}
The functions $b(t,x,\ep),\sigma(t,x,\ep)$ and $\gamma(t,x,z,\ep)$ 
are continuous in all their arguments and
$n^{\rm ae}$-time differentiable in $(x,\ep)$ with continuous derivatives.
Furthermore, 
there exists some positive constant $K$ such that\\
(i) for every $0\leq m \leq n^{\rm ae}$, $|\part_\ep^m b(t,0,\ep)|+|\part_\ep^m \sigma(t,0,\ep)|\leq K$
uniformly in $(t,\ep)\in [0,T]\times [0,1]$,\\
(ii) for every $1\leq n\leq n^{\rm ae} , 0\leq m\leq n^{\rm ae}$, 
$|\part_x^n\part_\ep^m b(t,x,\ep)|+|\part_x^n \part_\ep^m \sigma(t,x,\ep)|\leq K$
uniformly in $(t,x,\ep)\in[0,T]\times \mbb{R}^d \times [0,1]$, \\
(iii) for every $0\leq m\leq n^{\rm ae}$ and column $1\leq i\leq k$, $|\part_\ep^m \gamma_{\cdot,i}(t,0,z,\ep)|/\eta(z)\leq K$ uniformly in $(t,z,\ep)\in[0,T]\times \mbb{R}_0\times [0,1]$,\\
(iv) for every 
$1\leq n\leq n^{\rm ae} , 0\leq m\leq n^{\rm ae}$
and column $1\leq i\leq k$, $|\part_x^n\part_\ep^m \gamma_{\cdot,i}(t,x,z,\ep)|/\eta(z)\leq K$
uniformly in $(t,x,z,\ep)\in[0,T]\times \mbb{R}^d\times \mbb{R}_0\times[0,1]$.
\end{assumption}

\begin{assumption}
\label{assumption-Y} There exist some positive constants $K, q$ such that \\
(i) $\xi(x)$ is $n^{\rm ae}$-time differentiable in $x$ with continuous derivatives. 
Moreover, it has at most polynomial growth $|\part_x^n \xi(x)|\leq K(1+|x|^q)~x\in\mbb{R}^d$ for every $0\leq n\leq n^{\rm ae}$, \\
(ii) $|\rho_i(z)|\leq K \eta(z)$ for every $1\leq i\leq k$ and $z\in \mbb{R}_0$,\\
(iii) $f(t,x,y,z,u)$ is continuous in all its arguments and $n^{\rm ae}$-time  differentiable 
in $(x,y,z,u)$ with continuous derivatives. Moreover, every partial derivative only in $x$ has at most polynomial
growth $|\part_x^n f(t,x,y,z,u)|\leq K(1+|x|^q), 1\leq n\leq n^{\rm ae}$
as well as all the other partial derivatives are bounded by K,  uniformly in $(t,x,y,z,u)\in[0,T]\times \mbb{R}^d \times \mbb{R}^m\times \mbb{R}^{m\times l} \times \mbb{R}^m$, \\
(iv) $|f(t,x,0,0,0)|\leq K(1+|x|^{q})$ for every $x\in\mbb{R}^d$ uniformly in $t\in[0,T]$.
\end{assumption}

We define $(\part_x X_s^{t,x,\ep},s\in[t,T])$ as the solution of the SDE (if exists) given by a formal differentiation:
\bea
\label{eq-delx-X}
&&\part_x X_s^{t,x,\ep}=\int_t^s\part_x b(r,X_r^{t,x,\ep},\ep)\part_x X_r^{t,x,\ep}dr
+\int_t^s \part_x \sigma(r,X_r^{t,x,\ep},\ep)\part_x X_r^{t,x,\ep}dW_r\nn \\
&&\qquad\quad+\int_t^s \int_E \part_x \gamma(r,X_r^{t,x,\ep},z,\ep)\part_x X_r^{t,x,\ep}\wt{\mu}(dr,dz)~,
\eea
similarly for $(\part_\ep X_s^{t,x,\ep},s\in[t,T])$ and every higher order flow
$(\part_x^n\part_\ep^m X_s^{t,x,\ep},s\in[t,T])_{m,n\geq 0}$.
\begin{proposition}
\label{prop-delxe-X}
Under Assumption~\ref{assumption-X}, the SDE (\ref{eq-X}) 
has a unique solution $X^{t,x,\ep}\in\mbb{S}^p_d[t,T]~\forall p\geq 2$. 
Furthermore, for $0\leq n, m \leq n^{\rm ae}$, every $(n,m)$-time classical differentiation of 
$X^{t,x,\ep}$ in $(x,\ep)$
is well defined and given by $\part_x^n\part_\ep^m X^{t,x,\ep}\in \mbb{S}^p_{d^{n+1}}[t,T]~\forall p\geq 2$,
which is a unique solution  of the corresponding SDE defined by the formal differentiation of the coefficients
as (\ref{eq-delx-X}).
\begin{proof}
The existence of a unique solution $X^{t,x,\ep}\in\mbb{S}_d^p[t,T]~\forall p\geq 2$
is standard and can be proved by Lemma~\ref{ap-lemma-X-existence}.
Since every SDE is linear, it is not difficult to 
recursively show that the same conclusion  holds for every $\part^n_x\part_\ep^m X^{t,x,\ep}$.
The agreement with the classical differentiation can be proved by following 
the arguments in Theorem 3.1 of Ma \& Zhang (2002)~\cite{Ma-Zhang}.
In particular,  one can show  
\bea
\lim_{h\rightarrow 0} \mbb{E}||\nabla X^h-\part_x X^{t,x,\ep}||_{[t,T]}^2=0 \nn
\eea
where $\displaystyle \nabla X^h_s:=\frac{X_s^{t,x+h,\ep}-X_s^{t,x,\ep}}{h}$ ($d=1$ for simplicity)
and similar relations for every higher order derivatives in $(x,\ep)$.
\end{proof}
\end{proposition}

\begin{proposition}
\label{prop-Y-existence}
Under Assumptions~\ref{assumption-X} and \ref{assumption-Y}, the BSDE (\ref{eq-Y}) has a unique solution
$(Y^{t,x,\ep},Z^{t,x,\ep},\psi^{t,x,\ep})$ which belongs to $\mbb{S}^p_m[t,T]\times \mbb{H}^p_{m\times l}[t,T]
\times \mbb{H}^p_{m,\nu}[t,T]~\forall p\geq 2$.
Furthermore, it also satisfies
\bea
||\hat{\Theta}^{t,x,\ep}||_{\calk^p[t,T]}\leq C_p(1+|x|^{q})
\eea
for every $p\geq 2$.
\begin{proof}
The existence of a unique solution follows from Lemma~\ref{ap-lemma-Y-existence}. In addition, one has
\bea
||\hat{\Theta}^{t,x,\ep}||^p_{\calk^p[t,T]}\leq C_p\mbb{E}\left[|\xi(X_T^{t,x,\ep})|^p
+\Bigl(\int_t^T |f(s,X_s^{t,x,\ep},0,0,0)|ds\Bigr)^p\right] \nn
\eea
and hence one obtains the desired result by Lemma~\ref{ap-lemma-X-existence} and the 
assumption of polynomial growth of $\xi(x)$ and $f(\cdot,x,0,0,0)$.
\end{proof}
\end{proposition}

To lighten the notation, we use the following symbol to represent the collective arguments:
\bea
&&\Theta_r^{t,x,\ep}:=\Bigl(X_r^{t,x,\ep}, Y_r^{t,x,\ep},Z_r^{t,x,\ep},
\int_{\mbb{R}_0} \rho(z)\psi_r^{t,x,\ep}(z)\nu(dz)\Bigr)\nn \\
&&\hat{\Theta}_r^{t,x,\ep}:=\Bigl(Y_r^{t,x,\ep},Z_r^{t,x,\ep},
\int_{\mbb{R}_0} \rho(z)\psi_r^{t,x,\ep}(z)\nu(dz)\Bigr)\nn.
\eea
We also use $\part_\Theta:=(\part_x,\part_y,\part_z,\part_u)$,
$\part_{\hat{\Theta}}:=(\part_y,\part_z,\part_u)$ and similarly for their higher order derivatives.

\begin{remark}
Let us remark on the practical implications of the Assumptions \ref{assumption-X} and \ref{assumption-Y},
since some readers may find that the smoothness assumptions are too restrictive.
In Appendix~\ref{sec-smoothing}, we prove a smooth approximation theorem for FBSDEs 
which justifies Assumptions~\ref{assumption-X} and \ref{assumption-Y} whenever the standard Lipschitz conditions
are satisfied.

Since the financial problems relevant for BSDEs are inevitably non-linear,  
we are forced to consider in a portfolio level. Thus, $\xi$ and $f$ are likely to 
be given by complicated piecewise linear functions, which involve a large number of 
non-smooth points.
The first step we can do is to approximate these functions by 
smooth ones by introducing mollifiers or projecting onto Chebyshev polynomials, for example.
In the industry, this is quite common even for linear products such as a digital option
to make delta hedging feasible in practice. A small additional fee arising from a mollifier is charged to a client
as a hedging cost. It is also used for CVA evaluation by Henry-Labord\`ere (2012)~\cite{Henry}.
\end{remark}
\subsection{Representation theorem for BSDEs}
We define the Malliavin derivatives $D_{t,z}$ according to the conventions
used in Section 3 of Delong \& Imkeller (2010)~\cite{Delong-Imkeller} and Section 2.6
of Delong (2013)~\cite{Delong} (with $\sigma=1$).
See also Di Nunno et al (2009)~\cite{Nunno} for details and other applications.

According to their definition, if the random variable $H(\cdot,\omega_\mu)$
is differentiable in the sense of classical Malliavin's calculus for $\mbb{P}_\mu$-a.e. $\omega_\mu\in \Omega_\mu$,
then we have the relation
\be
D_{t,0}H(\omega_W,\omega_\mu)=D_tH(\cdot,\omega_\mu)(\omega_W)~, \nn
\ee
where $D_\cdot$ is the Malliavin's derivative with respect to the Wiener direction.
For the definition $D_{t,z}H$ with $z\neq 0$, the increment quotient operator~
is introduced
\be
\cali_{t,z}H(\omega_W,\omega_\mu):=\frac{H(\omega_W,\omega_\mu^{t,z})-H(\omega_W,\omega_\mu)}{z} \nn
\ee
where $\omega^{t,z}_\mu$ transforms a family $\omega_\mu=((t_1,z_1),(t_2,z_2),\cdots)\in\Omega_\mu$
into a new family $\omega_\mu^{t,z}((t,z),(t_1,z_1)$, $(t_2,z_2),\cdots)\in\Omega_\mu$.
This is defined for a one-dimensional Poisson random measure.
In the multi-dimensional case, $\cali_{t,z}H$ is extended to a $k$-dimensional vector in the obvious way.
It is known that when $\mbb{E}\Bigl[\int_0^T\int_E |\cali_{t,z}H|^2z^2\nu(dz)dt\Bigr]=
\mbb{E}\Bigl[\sum_{i=1}^k \int_0^T\int_{\mbb{R}_0} |\cali_{t,z_i}H|^2z_i^2\nu^i(dz_i)dt\Bigr]<\infty$, one has
$D_{t,z}H=\cali_{t,z}H$.

\begin{proposition}
Under Assumption~\ref{assumption-X}, the process $X^{t,x,\ep}$ is Malliavin differentiable.
Moreover, it satisfies
\bea
\sup_{(s,z)\in [0,T]\times \mbb{R}^k} \mbb{E}\Bigl[
\sup_{r\in[s,T]}|D_{s,z}X^{t,x,\ep}_r|^p\Bigr]<\infty \nn
\eea
for  $\forall p\geq 2$.
\begin{proof}
This is a modification of Theorem 4.1.2 in \cite{Delong} for our setting.
The existence of Malliavin derivative follows from Theorem 3 in Petrou (2008)~\cite{Petrou}.

According to \cite{Petrou}, for $z^i\neq 0$, one has
\bea
&&D_{s,z^i}X_r^{t,x,\ep}=\frac{\gamma^i(s,X_{s-}^{t,x,\ep},z^i,\ep)}{z^i}+\int_s^r D_{s,z^i}b(u,X_u^{t,x,\ep},\ep)du \nn \\
&&\qquad +\int_s^r D_{s,z^i}\sigma(u,X_u^{t,x,\ep},\ep)dW_u+\int_s^r \int_E 
D_{s,z^i}\gamma(u,X_{u-}^{t,x,\ep},z,\ep)\wt{\mu}(du,dz)
\label{eq-DX}
\eea
for $s\leq r$ and $D_{s,z^i}X_r^{t,x,\ep}=0$ otherwise.
Here, $\gamma^i$ denotes the $i$-th column vector and
\bea
&&D_{s,z^i}b(u,X_u^{t,x,\ep},\ep):=\frac{1}{z^i}\bigl[ b(u,X_u^{t,x,\ep}+z^iD_{s,z^i}X_u^{t,x,\ep},\ep)
-b(u,X_u^{t,x,\ep},\ep)\bigr] \nn
\eea
and similarly for the terms $(D_{s,z^i}\sigma(u,X_u^{t,x,\ep},\ep), D_{s,z^i}\gamma(u,X_{u-}^{t,x,\ep},z,\ep))$.
Due to the uniformly bounded derivative of $\part_x b,\part_x \sigma, \part_x \gamma/\eta$,
(\ref{eq-DX}) has the unique solution by Lemma~\ref{ap-lemma-X-existence}.
In addition,  applying the Burkholder-Davis-Gundy (BDG), Gronwall inequalities and Lemma~\ref{ap-lemma-BGJ},
one obtains
\bea
\mbb{E}||D_{s,z^i}X^{t,x,\ep}||_{[s,T]}^p\leq C_p
\Bigl(\Bigl|\frac{\gamma^i(s,0,z^i,\ep)}{z^i}\Bigr|^p+\mbb{E}||X^{t,x,\ep}||_T^p\Bigr)~. \nn
\eea
Thus, by Assumption~\ref{assumption-X} (iii), we obtain the desired result.
The arguments for the Wiener direction $(z=0)$ are similar.
\end{proof}
\end{proposition}

Next theorem is an adaptation of Theorem 3.5.1 in \cite{Delong} and Theorem C.1 in \cite{FT-qg-jumps}.
We suppress the superscripts $(t,x,\ep)$ denoting the initial data for simplicity.

\begin{theorem}
\label{theorem-representation}
Under Assumptions \ref{assumption-X} and \ref{assumption-Y}, \\
(a) There exists a unique solution $(Y^{s,0},Z^{s,0},\psi^{s,0})$ belongs to $\calk^p~\forall p\geq 2$ to the BSDE
\bea
Y_{u}^{s,0}=D_{s,0}\xi(X_T)+\int_u^T f^{s,0}(r)dr-\int_u^T Z_r^{s,0}dW_r
-\int_u^T\int_E \psi^{s,0}_r(z)\wt{\mu}(dr,dz)\nn
\eea
where
\bea
D_{s,0}\xi(X_T)&:=&\part_x \xi(X_T)D_{s,0}X_T\nn \\
f^{s,0}(r)&:=&\part_x f(r,\Theta_r)D_{s,0}X_r+\part_y f(r,\Theta_r)Y^{s,0}_r+\part_z f(r,\Theta_r)Z^{s,0}_r\nn \\
&+&\part_u f(r,\Theta_r)\int_{\mbb{R}_0} \rho(z)\psi_r^{s,0}(z)\nu(dz).\nn
\eea
(b) For $z^i\neq 0$, there exists a unique solution $(Y^{s,z^i},Z^{s,z^i},\psi^{s,z^i})$ belongs to 
$\calk^p~\forall p\geq 2$ to the BSDE
\bea
Y_u^{s,z^i}=D_{s,z^i}\xi(X_T)+\int_u^T f^{s,z^i}(r)dr-\int_u^T Z^{s,z^i}_rdW_r
-\int_u^T\int_E \psi^{s,z^i}_r(z)\wt{\mu}(dz,dr)\nn
\eea
where
\bea
D_{s,z^i}\xi(X_T)&:=&\frac{\xi(X_T+z^iD_{s,z^i}X_T)-\xi(X_T)}{z^i} \nn \\
f^{s,z^i}(r)&:=&\Bigl[f\Bigl(r,X_r+z^iD_{s,z^i}X_r,Y_r+z^iD_{s,z^i}Y_r,Z_r+z^iD_{s,z^i}Z_r \nn \\
&&\hspace{-12mm},\int_{\mbb{R}_0}\rho(e)\bigl[\psi_r(e)+z^iD_{s,z^i}\psi_r(e)]\nu(de)\Bigr)-f\Bigl(r,X_r,Y_r,Z_r,
\int_{\mbb{R}_0}\rho(e)\psi_r(e)\nu(de)\Bigr)
\Bigr]/z^i \nn
\eea
for every $1\leq i\leq k$.\\
(c)For $u<s\leq T$, set $(Y^{s,z}_u,Z^{s,z}_u,\psi^{s,z}_u)=0$ for $z\in\mbb{R}^k$ (i.e., including Wiener direction $z=0$). 
Then, $(Y,Z,\psi)$ is Malliavin differentiable and $(Y^{s,z},Z^{s,z},\psi^{s,z})$
is a version of $(D_{s,z}Y, D_{s,z}Z,D_{s,z}\psi)$.\\
(d)Set a deterministic function $u(t,x,\ep):=Y_t^{t,x,\ep}$ using the solution of the BSDE (\ref{eq-Y}).
If $u$ is continuous in $t$ and one-time continuously differentiable with respect to $x$, then 
\bea
\label{eq-Z-rep}
&&Z^{t,x,\ep}_s=\part_x u(s,X^{t,x,\ep}_{s-},\ep)\sigma(s, X^{t,x,\ep}_{s-},\ep) \\
&&\Bigl(\psi^{t,x,\ep}_s(z)\Bigr)^i_{1\leq i\leq k}=
\Bigl(u\bigl(s,X_{s-}^{t,x,\ep}+\gamma^i(s,X_{s-}^{t,x,\ep},z^i,\ep),\ep\bigr)
-u(s,X_{s-}^{t,x,\ep},\ep)\Bigr)_{1\leq i\leq k}
\label{eq-psi-rep}
\eea
for $t\leq s\leq T$ and $z=(z^i)_{1\leq i\leq k}\in \mbb{R}^k$.
\begin{proof}
(a) and (b) can be proved by Lemma~\ref{ap-lemma-Y-existence}, the boundedness of derivatives and 
the fact that $\Theta^{t,x,\ep}\in \mbb{S}^p\times \calk^p$ and $D_{s,z}X\in \mbb{S}^p$ for $\forall p\geq 2$. \\
(c) can be proved as a simple modification of Theorem 3.5.1 in \cite{Delong},
which is an extension of Proposition 5.3 in El Karoui et.al (1997)~\cite{ElKaroui}
to the jump case. The conditions written for $\omega$-dependent driver (assumptions (vii) and (viii) of \cite{Delong})
can be replaced by our assumption on $f$, which is Lipschitz with respect to $(y,z,u)$ and
has a polynomial growth in $x$. Note that we already know $X^{t,x,\ep},D_{s,z}X^{t,x,\ep}\in\mbb{S}^p~\forall p\geq 2$.
See also the arguments used in proof of Theorem 6.1 in \cite{FT-qg-jumps} for a Markovian setup.
(d) follows from Theorem 4.1.4 of \cite{Delong}. 
\end{proof}
\end{theorem}

\section{Asymptotic Expansion}
As the asymptotic expansion scheme, we want to obtain the Taylor expansion 
of the solution  $(X^{t,x,\ep},Y^{t,x,\ep},Z^{t,x,\ep}, \psi^{t,x,\ep})$ of the FBSDEs (\ref{eq-X})
and (\ref{eq-Y}) around $\ep=0$.
It is well-known that this is possible for the forward process $X^{t,x,\ep}$.
For the backward components $\hat{\Theta}^{t,x,\ep}$, we need to prove
the existence of classical derivative $\part_\ep^n \hat{\Theta}^{t,x,\ep}$ for every $0 \leq n\leq n_{\rm max}+1$
and then to obtain its estimate in an appropriate norm.
Since the BSDE corresponding to the classical derivative $\part_\ep^n \hat{\Theta}^{t,x,\ep}$
contains the terms proportional to  $\prod_{i=1}^j \part_\ep^{k_i}\hat{\Theta}^{t,x,\ep}$ with $\sum_{i=1}^j k_i=n$
in its driver, the estimates of $(\part_\ep^{k_i} Z^{t,x,\ep}, \part_\ep^{k_i} \psi^{t,x,\ep})_{i=1}^j$
with respect to the norm $\mbb{H}^p\times \mbb{H}^{p}_{\nu}~\forall p\geq 2$ are not enough 
to guarantee the well-posedness of the relevant BSDE.

In the following, we shall solve this issue by showing $(\part^{k_i}_\ep Z^{t,x,\ep}, \part_\ep^{k_i} \psi^{t,x,\ep})$
actually belongs to $\mbb{S}^p\times \mbb{S}^p~\forall p\geq 2$ instead of $\mbb{H}^p\times \mbb{H}^p_{\nu}~\forall p\geq 2$.
This is done by recursively applying the representation theorem and the polynomial growth property of the solutions with respect to $x$.
In order to use the result in Theorem~\ref{theorem-representation} (d), we have to start
from studying the classical derivatives of the BSDE (\ref{eq-Y}) with respect to $x$.

\subsection{Classical derivatives of BSDEs}

\begin{lemma}
\label{lemma-delxY}
Under Assumptions \ref{assumption-X} and \ref{assumption-Y}, 
$\hat{\Theta}^{t,x,\ep}$ is classically differentiable in $x$,  and it is given by 
$\part_x\hat{\Theta}^{t,x,\ep}$ defined as the unique solution of 
the BSDE with formal differentiation:
\bea
&&\part_x Y^{t,x,\ep}_s=\part_x \xi(X_T^{t,x,\ep})\part_x X_T^{t,x,\ep}
+\int_s^T \part_\Theta f(r,\Theta_r^{t,x,\ep})\part_x \Theta_r^{t,x,\ep}dr\nn \\
&&\quad -\int_s^T \part_x Z_r^{t,x,\ep}dW_r-\int_s^T \int_E \part_x \psi_r^{t,x,\ep}(z)\wt{\mu}(dr,dz)
\label{eq-delxY}
\eea
and $\part_x \hat{\Theta}^{t,x,\ep}\in \calk^p[t,T]$ satisfying 
\be
||\part_x\hat{\Theta}^{t,x,\ep}||_{\calk^p[t,T]}\leq C_p (1+|x|^{q}) \nn
\ee
for  $\forall p\geq 2$.
\begin{proof}
The existence and uniqueness can be easily shown by Lemma~\ref{ap-lemma-Y-existence}.
Note that the BSDE (\ref{eq-delxY}) is  linear with bounded Lipschitz constants and satisfies
\bea
&&||\part_x\hat{\Theta}^{t,x,\ep}||^p_{\calk^p[t,T]}\leq
C_p\mbb{E}\Bigl[|\part_x \xi(X_T^{t,x,\ep})|^p |\part_x X_T^{t,x,\ep}|^p+
\Bigl(\int_t^T |\part_x f(r,\Theta_r^{t,x,\ep})||\part_x X_r^{t,x,\ep}|dr\Bigr)^p\Bigr] \nn \\
&&\leq C_p||\part_x X^{t,x,\ep}||^p_{\mbb{S}^{2p}[t,T]}\Bigl\{ \Bigl(\mbb{E}|\part_x \xi(X_T^{t,x,\ep})|^{2p}\Bigr)^{1/2}
+\Bigl(\mbb{E}\Bigl(\int_t^T|\part_x f(r,X_r^{t,x,\ep},0)|dr\Bigr)^{2p}\Bigr)^{1/2}\nn \\
&&+||\hat{\Theta}^{t,x,\ep}||^p_{\calk^{2p}[t,T]}\Bigr\}
\leq C_p(1+|x|^{pq}) \nn
\eea
for  $\forall p\geq 2$.
With a simple modification of Theorem 3.1 of \cite{Ma-Zhang}, one can also show that
\bea
\lim_{h\rightarrow 0} ||\nabla^h \hat{\Theta}^{t,x,\ep}-\part_x \hat{\Theta}^{t,x,\ep}||^2_{\calk^2[t,T]}=0 \nn
\eea
where $\displaystyle \nabla^h \hat{\Theta}^{t,x,\ep}:=\frac{\hat{\Theta}^{t,x+h,\ep}-\hat{\Theta}^{t,x,\ep}}{h}$
with $h\neq 0$ (for each direction). This gives the agreement with the classical differentiation.
\end{proof}
\end{lemma}

\begin{corollary}
\label{corollary-Theta}
Under Assumptions \ref{assumption-X} and \ref{assumption-Y}, 
there exists $\part_x u(t,x,\ep)$ which is continuous in $(t,x)$ and has at most a polynomial growth in $x$ 
uniformly in $(t,\ep)\in[0,T]\times [0,1]$.
Furthermore, $Z^{t,x,\ep}$ and $\int_{\mbb{R}_0} \rho(z) \psi^{t,x,\ep}(z)\nu(dz)$ belong to
$\mbb{S}^p[t,T]~\forall p\geq 2$.
\begin{proof}
Note that $\part_x u(t,x,\ep)=\part_x Y^{t,x,\ep}_t$ and there exists some constant $C>0$ such that
\be
|\part_x u(t,x,\ep)|\leq ||\part_x \hat{\Theta}^{t,x,\ep}||_{\calk^p{[t,T]}}\leq C(1+|x|^{q})\nn
\ee
for every $x\in\mbb{R}^d$ uniformly in $t\in[0,T]$ by Lemma~\ref{lemma-delxY}.
The continuity of $\part_x u(t,x,\ep)$ in $(t,x)$ can be shown in the same way as \cite{Ma-Zhang}
using the continuity of $X^{t,x,\ep}$ in $(t,x)$, which can be seen in Lemma~\ref{ap-lemma-X-existence}. 
Then, from the representation given in (\ref{eq-Z-rep}), (\ref{eq-psi-rep}) and the above result,
one sees
\bea
|Z^{t,x,\ep}_s|+\Bigl|\int_E \rho(z)\psi_s^{t,x,\ep}(z)\nu(dz)\Bigr|\leq C(1+|X_{s-}^{t,x,\ep}|^{q+1}) \nn
\eea
which gives the desired result $\hat{\Theta}^{t,x,\ep}\in\mbb{S}^p[t,T]^{\otimes 3}$ for any $p\geq 2$.
\end{proof}
\end{corollary}

\begin{proposition}
\label{prop-delxnY}
Under Assumptions \ref{assumption-X} and \ref{assumption-Y}, the classical derivative
$\part_x^n \hat{\Theta}^{t,x,\ep}$ exists for every $0\leq n\leq n^{\rm ae}$
with $\part^{n}_x \hat{\Theta}^{t,x,\ep}\in \calk^p[t,T]~\forall p\geq 2$ and is given by the solution of the following BSDE:
\bea
\part_x^n Y_s^{t,x,\ep}&=&\xi_n+\int_s^T\Bigl\{H_{n,r}+\part_\Theta f(r,\Theta_r^{t,x,\ep})\part_x^n \Theta_r^{t,x,\ep}
\Bigr\}dr\nn \\
&&-\int_s^T \part_x^nZ_r^{t,x,\ep}dW_r-\int_s^T\int_E \part_x^n \psi^{t,x,\ep}_r(z)\wt{\mu}(dr,dz)
\label{eq-delxnY}
\eea
where
\bea
&&\xi_n:=n!\sum_{k=1}^n \sum_{\beta_1+\cdots+\beta_k=n,\beta_i\geq 1}\frac{1}{k!}\part_x^k \xi(X_T^{t,x,\ep})
\prod_{j=1}^k\frac{1}{\beta_j!}\part_x^{\beta_j}X_T^{t,x,\ep},\nn\\
&&H_{n,r}:=n!\sum_{k=2}^n \sum_{\beta_1+\cdots+\beta_k=n,\beta_i\geq 1}\sum_{i_x=0}^k
\sum_{i_y=0}^{k-i_x}\sum_{i_z=0}^{k-i_x-i_y}
\frac{\part_x^{i_x}\part_y^{i_y}\part_z^{i_z}\part_u^{k-i_x-i_y-i_z}f(r,\Theta_r^{t,x,\ep})}{i_x!
i_y!i_z!(k-i_x-i_y-i_z)!}\nn \\
&&\quad \times \prod_{j_x=1}^{i_x}\frac{1}{\beta_{j_x}!}\part_x^{\beta_{j_x}}X_r^{t,x,\ep}
\prod_{j_y=i_x+1}^{i_x+i_y}\frac{1}{\beta_{j_y}!}\part_x^{\beta_{j_y}}Y_r^{t,x,\ep}
\prod_{j_z=i_x+i_y+1}^{i_x+i_y+i_z}\frac{1}{\beta_{j_z}!}\part_x^{\beta_{j_z}}Z_r^{t,x,\ep}\nn \\
&&\quad \times \prod_{j_u=i_x+i_y+i_z+1}^{k}\frac{1}{\beta_{j_u}!}
\int_{\mbb{R}_0} \rho(z)\part_x^{\beta_{j_u}}\psi^{t,x,\ep}_r(z)\nu(dz)~.\nn
\eea
Moreover, for every $0\leq n\leq n_{\rm max}+1$, $\part_x^n \hat{\Theta}^{t,x,\ep}\in \mbb{S}^p[t,T]^{\otimes 3}~ \forall p\geq 2$.
\begin{proof}
We can prove recursively by the arguments used in Proposition \ref{prop-Y-existence}, 
Lemma \ref{lemma-delxY} and Corollary \ref{corollary-Theta}.
We already know that $\hat{\Theta}^{t,x,\ep}\in\mbb{S}^p[t,T]^{\otimes 3}$ 
and $\part_x \hat{\Theta}^{t,x,\ep}\in\calk^p[t,T]$ for any $p\geq 2$.
The BSDE for $\part_x^2 \hat{\Theta}^{t,x,\ep}$ has bounded Lipschitz constants
and $H_{2,r}$ is at most quadratic in $(\part_x\hat{\Theta}^{t,x,\ep}_r)$.
From the fact that $\xi(x),f(\cdot,x,0)$ have at most a polynomial growth in $x$
and that $(\part_x^m X^{t,x,\ep})_{0\leq m\leq n^{\rm ae}}, \hat{\Theta}^{t,x,\ep}\in \mbb{S}^p[t,T]~\forall p\geq 2$,
one can prove the existence of the unique solution $\part_x^2 \hat{\Theta}^{t,x,\ep}\in\calk^p[t,T]~\forall p\geq 2$
by Lemma~\ref{ap-lemma-Y-existence}.
Furthermore, one can show as in Lemma \ref{lemma-delxY} that
$||\part_x^2\hat{\Theta}^{t,x,\ep}||_{\calk^p[t,T]}$ has at most polynomial growth in $x$.
By following the arguments of Theorem 3.1 of \cite{Ma-Zhang}, one sees this agrees with 
the classical differentiation in the sense of Lemma \ref{lemma-delxY}.
This in turn shows the existence  $\part_x^2 u(t,x,\ep)=\part_x^2 Y_t^{t,x,\ep}$
and also the fact that $\part_x^2 u(t,x,\ep)$ has at most a polynomial growth in $x$.
This implies that,  together with Assumption~\ref{assumption-X} and the representation theorem (\ref{eq-Z-rep}) (\ref{eq-psi-rep}),
$\part_x Z^{t,x,\ep}$ and $\int_{\mbb{R}_0} \rho(z)\part_x \psi^{t,x,\ep}(z)\nu(dz)$ are in $\mbb{S}^p[t,T]~\forall p\geq 2$.
Thus, we get $\part_x \hat{\Theta}^{t,x,\ep}\in\mbb{S}^p[t,T]^{\otimes 3}~\forall p\geq 2$.

In the same manner, if we assume that $\Bigl(\part_x^i \hat{\Theta}^{t,x,\ep}\Bigr)_{i\leq n}\in \mbb{S}^p[t,T]^{\otimes 3}$
and that $\part_x^{n+1}\hat{\Theta}^{t,x,\ep}\in\calk^p[t,T]$ for $\forall p\geq 2$ with the $\calk^p$-norm 
at most a polynomial growth in $x$ ,
then one can show that the existence of the unique solution $\part_x^{n+2}\hat{\Theta}^{t,x,\ep}\in\calk^p[t,T]$
with the norm at most a polynomial growth in $x$ by Lemma~\ref{ap-lemma-Y-existence}.
It  then implies from the representation theorem that 
$\part_x^{n+1}\hat{\Theta}^{t,x,\ep}\in \mbb{S}^p[t,T]^{\otimes 3}~\forall p\geq 2$.
By repeating the procedures, one obtains the desired result.
\end{proof}
\end{proposition}

\subsection{Asymptotic expansion}
We are now going to prove $\part_\ep^n \hat{\Theta}^{t,x,\ep}\in\mbb{S}^p[t,T]^{\otimes 3}~\forall p\geq 2$
for every $0\leq n\leq n_{\rm max}+1$. Although the strategy is similar to the previous section, we actually have to study
the properties of $\bigl(\part_x^m\part_\ep^n \hat{\Theta}^{t,x,\ep}\bigr)$ since $\ep$ affects $u(s,X_{s-}^{t,x,\ep},\ep)$
not only through its explicit dependence but also through $X^{t,x,\ep}$ indirectly.

\begin{lemma}
\label{lemma-deleY}
Under Assumptions \ref{assumption-X} and \ref{assumption-Y}, $\hat{\Theta}^{t,x,\ep}$
is classically differentiable  in $\ep$, and it is given by $\part_\ep \hat{\Theta}^{t,x,\ep}$
defined as the unique solution of the BSDE with formal differentiation:
\bea
&&\part_\ep Y_s^{t,x,\ep}=\part_x \xi(X_T^{t,x,\ep})\part_\ep X_T^{t,x,\ep}
+\int_s^T \part_\Theta f(r,\Theta_r^{t,x,\ep})\part_\ep \Theta_r^{t,x,\ep}dr\nn \\
&&\quad-\int_s^T \part_\ep Z_r^{t,x,\ep}dW_r-\int_s^T\int_E \part_\ep\psi^{t,x,\ep}_r\wt{\mu}(dr,dz)~.\nn
\eea
One has $\part_\ep \hat{\Theta}^{t,x,\ep}\in\calk^p[t,T]$ satisfying
\be
||\part_\ep \hat{\Theta}^{t,x,\ep}||_{\calk^p[t,T]}\leq C_p(1+|x|^{q}) \nn
\ee
for any $\forall p\geq 2$. 
\begin{proof}
The proof can be done similarly as in Lemma~\ref{lemma-delxY}.
\end{proof}
\end{lemma}

We now get the following result.
\begin{proposition}
\label{prop-delenY}
Under Assumptions \ref{assumption-X} and \ref{assumption-Y}, the classical derivative $\part_\ep^n \hat{\Theta}^{t,x,\ep}$ 
exists for every $0 \leq n\leq n^{\rm ae}$ with $\part_\ep^n \hat{\Theta}^{t,x,\ep}\in\calk^p[t,T]~\forall p\geq 2$ 
and is given by the unique solution of the following BSDE:
\bea
&&\part^n_\ep Y_s^{t,x,\ep}=\wt{\xi}_n+\int_s^T \Bigl\{\wt{H}_{n,r}+
\part_\Theta f(r,\Theta_r^{t,x,\ep})\part^n_\ep \Theta^{t,x,\ep}_r\Bigr\}dr\nn \\
&&\qquad-\int_s^T \part_\ep^n Z_r^{t,x,\ep}dW_r-\int_s^T\int_E \part_\ep^n \psi_r^{t,x,\ep}\wt{\mu}(dr,dz)~.\nn 
\eea
Here, $\wt{\xi}_n$ and $\wt{H}_{n,r}$ are
given by the expressions of $\xi_n$ and $H_{n,r}$ in Proposition~\ref{prop-delxnY}
with $\part_x^{\beta_{j_\Theta}}$ replaced by $\part_\ep^{\beta_{j_\Theta}}$.
Moreover, for every $0\leq n\leq n_{\rm max}+1$, $\part_\ep^n\hat{\Theta}^{t,x,\ep}\in\mbb{S}^p[t,T]^{\otimes 3}~\forall p\geq 2$.
\begin{proof}
We start from the result of Lemma~\ref{lemma-deleY}, which implies $\part_\ep u(t,x,\ep)$
has at most polynomial growth in $x$.
Using the fact that $\part_\ep \Theta^{t,x,\ep}\in\mbb{S}^p[t,T]\times \calk^p[t,T]$
and $\part_x \Theta^{t,x,\ep}\in \mbb{S}^p[t,T]^{\otimes 4}$, 
one can show that $\part_x \part_\ep \hat{\Theta}^{t,x,\ep}$ exists and satisfies 
$\part_x \part_\ep \hat{\Theta}^{t,x,\ep}\in \calk^p[t,T]$ for $\forall p\geq 2$
as in Lemma~\ref{lemma-delxY}. The corresponding norm has at most
polynomial growth in $x$ and so is $\part_x\part_\ep u(t,x,\ep)$.
This implies, together with the representations (\ref{eq-Z-rep}) and (\ref{eq-psi-rep}), 
that $\part_\ep\hat{\Theta}^{t,x,\ep}\in \mbb{S}^p[t,T]$ for $\forall p\geq 2$.

As in Proposition~\ref{prop-delxnY}, one can recursively prove that the classical derivative
$\part_x^n\part_\ep \hat{\Theta}^{t,x,\ep}$ exists and belongs to $\calk^p[t,T]~\forall p\geq 2$
for every $0\leq n\leq n^{\rm ae}$ and moreover that it belongs to $\mbb{S}^p[t,T]^{\otimes 3}~\forall p\geq 2$
for every $0\leq n\leq n_{\rm max}+1$ by induction.
Then, by Lemma~\ref{ap-lemma-Y-existence}, it is straightforward to check $\part_x^n \part_\ep^2 \hat{\Theta}^{t,x,\ep}$
exists and belongs to $\calk^p[t,T]~\forall p\geq 2$ for $0\leq n\leq n^{\rm ae}$.
By the representation theorem, it then implies $\part_x^n \part_\ep^2 \hat{\Theta}^{t,x,\ep}$ 
in fact belongs to $\in \mbb{S}^p[t,T]~\forall p\geq 2$ for $0\leq n\leq n_{\rm max}+1$.
By repeating the same procedures,  one can show that, for every $0\leq n,m\leq n_{\rm max}+1$, 
$\part_x^n\part_\ep^m \hat{\Theta}^{t,x,\ep}$ exists and belongs to $\mbb{S}^p[t,T]^{\otimes 3}~\forall p\geq 2$.  Thus the claims of the proposition are proved.
\end{proof}
\end{proposition}

We have shown that $\Theta^{t,x,\ep}$ is $n^{\rm ae}$-time classically differentiable with respect to $(x,\ep)$
and, in particular for $n\leq n_{\rm max}+1$, $\part_\ep^n \Theta^{t,x,\ep}\in\mbb{S}^p[t,T]^{\otimes 4}~\forall p\geq 2$. 
Let us define for $s\in[t,T]$ and $0\leq n\leq n_{\rm max}$ that
\bea
\Theta_s^{[n]}:=\frac{1}{n!}\part_\ep^n \Theta^{t,x,\ep}_s\Bigr|_{\ep=0}.\nn
\eea
Using the differentiability and the Taylor formula, one has for any $1\leq N\leq n_{\rm max}$
\bea
\Theta_s^{t,x,\ep}=\Theta_s^{[0]}+\sum_{n=1}^{N} \ep^n\Theta_s^{[n]}+\frac{\ep^{N+1}}{N!}
\int_0^1 (1-u)^{N} \bigl(\part_\alpha^{N+1}\Theta_s^{t,x,\alpha}\bigr)\Bigr|_{\alpha=u\ep}du~.
\label{eq-AE-Theta}
\eea
As we shall see later, each $\Theta^{[m]}, m\in\{1,2,\cdots, n_{\rm max}\}$
can be evaluated by solving the system of linear ODEs.
Although $\Theta^{[0]}$ requires to solve a non-linear ODE as an exception,
the existence of the bounded solution is guaranteed under the Assumptions \ref{assumption-X} and \ref{assumption-Y}.

The next theorem is the  main result of the paper which gives the error estimate for the approximation 
of $\Theta^{t,x,\ep}$ by the series of $\Theta^{[m]}, m\in\{0,1,\cdots,n_{\rm max}\}$.
\begin{theorem}
\label{theorem-AE}
Under Assumptions~\ref{assumption-X} and \ref{assumption-Y}, 
the asymptotic expansion of the forward-backward SDEs (\ref{eq-X}) and (\ref{eq-Y}) 
is given by (\ref{eq-AE-Theta}) for every $1\leq N\leq n_{\rm max}$ and satisfies, with some positive constant $C_p$, that
\bea
\left|\left|\Theta^{t,x,\ep}-\Bigl(\Theta^{[0]}+\sum_{n=1}^{N} \ep^n \Theta^{[n]}\Bigr)\right|\right|_{\mbb{S}^p[t,T]}
\leq \ep^{N+1}C_p~.
\label{eq-error-asymp}
\eea
\begin{proof}
This immediately follows from the fact that $\part^{N+1}_{\ep}\Theta^{t,x,\ep}$ is in $\mbb{S}^p[t,T]~\forall p\geq 2$ and continuous 
with respect to $\ep$ by Propositions \ref{prop-delxe-X} and \ref{prop-delenY}.
\end{proof}
\end{theorem}

\subsection{State-dependent jump intensity}
When $\nu$ is a finite measure $\nu(E)<\infty$, all the 
previous results hold true with slightly weaker assumptions with $\eta, \rho\equiv 1$
in Assumptions \ref{assumption-X} and \ref{assumption-Y}.
In practical applications, however, there are many cases where we want to make 
the jump intensity state dependent.
In this section, we solve this problem when the intensity is bounded.

In particular, we consider the forward-backward SDEs (\ref{eq-X}) and (\ref{eq-Y}) but with
the compensated random measure $\wt{\mu}(dr,dz)$ given by, for $1\leq i \leq k$,
\bea
\wt{\mu}^i(dr,dz)=\mu^i(dr,dz)-\lambda^i(r,X_r^{t,x,\ep})\nu^i(dz)dr \nn
\eea
where $\nu^i$ is normalized as $\nu^i(\mbb{R}_0)=1$ and
$\lambda^i:[0,T]\times \mbb{R}^d\rightarrow \mbb{R}$. One can see that the random measure is not Poissonian 
any more and depends implicitly on $\ep$ through its intensity.
\begin{assumption}
\label{assumption-intensity}
For every $1\leq i\leq k$, $\nu^i(\mbb{R}_0)=1$ and there exist some positive constants $K, c_1, c_2$ such that\\
(i) $\lambda^i(t,x)$ is continuous in $(t,x)$,
$n^{\rm ae}$-time differentiable in $x$ with continuous derivatives satisfying 
$|\part_x^n \lambda^i(t,x)|\leq K$ for every $1\leq n\leq n^{\rm ae}$ uniformly in $(t,x)\in[0,T]\times \mbb{R}^d$,\\
(ii) $0<c_1\leq \lambda^i(t,x)\leq c_2 $ uniformly in $(t,x)\in[0,T]\times \mbb{R}^d$, \\
(iii) $|\part_\ep^m \gamma_{\cdot,i}(t,x,z,\ep)|\leq K$ for every $1\leq m \leq n^{\rm ae}$
uniformly in $(t,x,z,\ep)\in[0,T]\times \mbb{R}^d \times \mbb{R}_0 \times [0,1]$.
\end{assumption}

\begin{lemma}
\label{lemma-Q}
Under Assumption~\ref{assumption-intensity}, one can define an equivalent probability measure $\mbb{Q}$ by, for $s\in[t,T]$, 
\bea
\frac{d\mbb{Q}}{d\mbb{P}}\Bigr|_{\calf_s}=M_s \nn
\eea
where $M$ is a strictly positive $\mbb{P}$-martingale given by
\bea
M_s=1+\sum_{i=1}^k \int_t^s M_{r-}\Bigl(\frac{c_2}{\lambda^i(r,X_{r-}^{t,x,\ep})}-1\Bigr)
\wt{\mu}^i(dr,\mbb{R}_0)~. \nn
\eea
Under the new measure $\mbb{Q}$, the compensated random measure becomes
\be
\wt{\mu}^{\mbb{Q}}(dr,dz)=\mu(dr,dz)-c_2\nu(dz)dt\nn
\ee
and hence $\mu$ is Poissonian. Moreover, for $\forall s\in[t,T]$, 
\be
M_s\geq \exp\bigl(-(c_2-c_1)k(T-t)\bigr)~.\nn
\ee
\begin{proof}
By Kazamaki~(1979)~\cite{Kazamaki}, it is known that if $X$ is a BMO martingale satisfying
$\Del X_t\geq -1+\del$ a.s. for all $t\in[0,T]$ with some strictly positive constant $\del>0$,
then Dol\'eans-Dade exponential $\cale(X)$ is uniformly integrable.
One can easily confirm that this condition is satisfied for a martingale
\be
\int^\cdot \Bigl(c_2/\lambda(s,X_s^{t,x,\ep})-1\Bigr)\wt{\mu}(ds,\mbb{R}_0)~. \nn
\ee
Thus the given measure change is well-defined and
the first claim follows from Theorem~41 in Chapter 3 of \cite{Protter}.
The second claim directly follows from the explicit expression
\bea
&&M_s=\prod_{i=1}^k\left\{
\prod_{0<r\leq s}\Bigl(\frac{c_2}{\lambda^i(r,X_{r-}^{t,x,\ep})}\Bigr)^{\Del \mu^i(r,\mbb{R}_0)}\exp\Bigl(
-\int_t^s (c_2-\lambda^i(r,X_{r-}^{t,x,\ep}))dr\Bigr)\right\}\nn \\
&&\quad \geq  \exp\Bigl(-\int_t^s k(c_2-c_1)dr\Bigr)~. \nn
\eea
\end{proof}
\end{lemma}

Under the measure $\mbb{Q}$, we have 
\bea
\label{eq-X-intensity}
X_s^{t,x,\ep}&=&x+\int_t^s \wt{b}(r,X_{r}^{t,x,\ep},\ep)dr+\int_t^s \sigma(r,X_r^{t,x,\ep},\ep)dW_r \nn \\
&&+\int_t^s\int_E \gamma(r,X_{r-}^{t,x,\ep},z,\ep)\wt{\mu}^{\mbb{Q}}(dr,dz) 
\eea
\bea
Y_s^{t,x,\ep}&=&\xi(X_T^{t,x,\ep})+\int_s^T \wt{f}\Bigl(r,X_r^{t,x,\ep},Y_r^{t,x,\ep},Z_r^{t,x,\ep},
\int_{\mbb{R}_0} \psi_r^{t,x,\ep}(z)\nu(dz)\Bigr)dr \nn \\
&&-\int_s^T Z_r^{t,x,\ep}dW_r-\int_s^T \int_E \psi_r^{t,x,\ep}(z)\wt{\mu}^{\mbb{Q}}(dr,dz)
\label{eq-Y-intensity}
\eea
where 
\bea
&&\wt{b}(s,x,\ep)=b(s,x,\ep)+\sum_{i=1}^k(c_2-\lambda^i(s,x))\int_{\mbb{R}_0}\gamma^i(s,x,z^i,\ep)\nu(dz^i)\nn \\
&&\wt{f}(s,x,y,z,u)=f(s,x,y,z,u)-\sum_{i=1}^k(c_2-\lambda^i(s,x))u^i. \nn
\eea
\begin{theorem}
Under Assumptions~\ref{assumption-X}, \ref{assumption-Y} with $\rho$ and $\eta$ replaced by $1$,
and Assumption \ref{assumption-intensity},
the solution $\Theta^{t,x,\ep}$ of the forward-backward SDEs (\ref{eq-X}) and (\ref{eq-Y})
allows the asymptotic expansion with respect to $\ep$ and satisfies the same error estimate (\ref{eq-error-asymp})
in the original measure $\mbb{P}$.
\begin{proof}
Assumption \ref{assumption-intensity} makes $(\wt{b},\wt{f})$ once again satisfy 
Assumptions \ref{assumption-X} and \ref{assumption-Y} with  $\rho,\eta$ replaced by $1$.
Therefore,  all the results in the previous sections hold true under the measure $\mbb{Q}$
to the equivalent FBSDEs (\ref{eq-X-intensity}) and (\ref{eq-Y-intensity}).
In particular this implies from Lemma~\ref{lemma-Q} that,  with some positive constant $C_p$,
\bea
\ep^{p(N+1)}C_p&\geq& \mbb{E}^{\mbb{Q}}\left[\sup_{s\in[t,T]}\Bigl|\Theta^{t,x,\ep}_s-
\Bigl(\Theta^{[0]}_s+\sum_{n=1}^N\ep^n \Theta^{[n]}_s\Bigr)\Bigr|^p\right]\nn \\
&=&\mbb{E}\left[M_T\sup_{s\in[t,T]}\Bigl|\Theta^{t,x,\ep}_s-
\Bigl(\Theta^{[0]}_s+\sum_{n=1}^N\ep^n \Theta^{[n]}_s\Bigr)\Bigr|^p\right]\nn \\
&\geq& \exp\bigl(-k(c_2-c_1)(T-t)\bigr)\mbb{E}\left[\sup_{s\in[t,T]}\Bigl|\Theta^{t,x,\ep}_s-
\Bigl(\Theta^{[0]}_s+\sum_{n=1}^N\ep^n \Theta^{[n]}_s\Bigr)\Bigr|^p\right]\nn~.
\eea
This proves the claim.
\end{proof}
\end{theorem}

\section{Implementation of the asymptotic expansion}
\subsection{Evaluation scheme}
\label{sec-implementation}
In this section, we explain how to calculate $\Theta^{[n]}$, $n\in\{0,1,\cdots, n_{\rm max}\}$ (semi)-analytically.
As we shall see, if we introduce $\ep$ in a specific way to the 
forward SDE (\ref{eq-X}), then the grading structure introduced by the asymptotic expansion 
yields a very simple scheme requiring only a system of linear ODEs to be solved
with only one exception at the zero-th order.
It is also remarkable that one can directly approximate not only $(Y^{t,x}, Z^{t,x})$
but also the $\mbb{L}^2(E;\nu)$-valued
process $\psi^{t,x}(\cdot)$. This looks almost infeasible for the standard regression-based simulation scheme.
\\

Let us put the initial time as $t=0$, and take $(m=d=l=1)$ for {\it notational simplicity}.
The extension to higher dimensional setups is straightforward for which 
one only needs proper indexing of each variable.
Let us adopt a following parametrization of $X$ with $\ep$
which obviously leads to {\it small-variance} expansion;
\bea
X_s^{\ep}=x+\int_0^s b(r,X_r^{\ep},\ep)dr+\int_0^s \ep \sigma(r,X_r^\ep)dW_r
+\int_0^s\int_{\mbb{R}_0}\ep \gamma(r,X_{r-}^{\ep},z)\wt{\mu}(dr,dz)~, \nn
\eea
where we omit the superscript denoting the initial data $(0,x)$.
One can see that the process $X^{\ep}$ becomes deterministic when $\ep\rightarrow 0$.
Similar to the standard applications~\cite{T-review}, this parameterization 
is crucial to obtain semi-analytic approximations.
We make Assumptions~\ref{assumption-X} and \ref{assumption-Y}
(or those replaced by $\rho=\eta=1$ and Assumption~\ref{assumption-intensity}) the standing assumptions for this section.
\begin{lemma}
\label{lemma-0th}
The zero-th order solution $\bigl(\Theta^{[0]}_s,s\in[0,T]\bigr)$ is given by
\bea
&&X_s^{[0]}=x+\int_0^s b(r,X_r^{[0]},0)dr \nn\\
\label{eq-Y-0th}
&&Y_s^{[0]}=\xi(X_T^{[0]})+\int_s^Tf(r,X_r^{[0]},Y_r^{[0]},0,0)dr  \\
&&Z^{[0]}=\psi^{[0]}(\cdot)\equiv 0~. \nn
\eea
which is continuous, deterministic and bounded.
\begin{proof}
Thanks to the Lipschitz continuity of $b, f$ with respect to $x, y$ respectively,
the claim can be proved by the standard results for the ODEs.
\end{proof}
\end{lemma}

Let us introduce some notations. We denote, for $s\in[0,T]$,
\bea
&&b^{[0]}(s):=b(s,X_s^{[0]},0),\quad \sigma^{[0]}(s):=\sigma(s,X_s^{[0]}),\quad
\gamma^{[0]}(s,z):= \gamma(s,X_s^{[0]},z)\nn \\
&&\xi^{[0]}:=\xi(X_T^{[0]}), \quad f^{[0]}(s):=f(s,X_s^{[0]},Y_s^{[0]},0,0), \nn \\
&&\Gamma^{[0]}(s):=\int_{\mbb{R}_0}\rho(z)\gamma^{[0]}(s,z)\nu(dz)~.\nn
\eea
As for derivatives, we denote for example
\bea
&&\part_x b^{[0]}(s):=\part_x b(s,x,0)\Bigr|_{x=X_s^{[0]}}, \quad \part_\ep b^{[0]}(s)=\part_\ep b(s,X_s^{[0]},\ep)\Bigr|_{\ep=0} \nn \\
&&\part_x \Gamma^{[0]}(s):=\int_{\mbb{R}_0}\rho(z)\part_x \gamma(s,x,z)\Bigr|_{x=X_s^{[0]}}\nu(dz)\nn
\eea
and the other terms in the obvious way.

For the first order of the expansion, we have to solve
\bea
\label{eq-X-1}
&&X_s^{[1]}=\int_0^s \bigl[\part_\ep b^{[0]}(r)+\part_x b^{[0]}(r)X_r^{[1]}\bigr]dr+\int_0^s\sigma^{[0]}(r)dW_r+\int_0^s \int_{\mbb{R}_0}\gamma^{[0]}(r,z)\wt{\mu}(dr,dz), \nn \\ \\
&&Y_s^{[1]}=\part_x \xi^{[0]}X_T^{[1]}+\int_s^T \part_\Theta f^{[0]}(r)\Theta_r^{[1]}dr
-\int_s^T Z_r^{[1]}dW_r-\int_s^T \int_{\mbb{R}_0}\psi_r^{[1]}(z)\wt{\mu}(dr,dz)~.\nn \\
\label{eq-Y-1}
\eea

\begin{lemma}
\label{lemma-ae-1}
There exists a unique solution $\Theta^{[1]}$ to (\ref{eq-X-1}) and (\ref{eq-Y-1}) which 
belongs to $\mbb{S}^p[0,T]^{\otimes 4}~\forall p\geq 2$. $\hat{\Theta}^{[1]}$ is given by,
for $s\in[0,T]$ and $z\in\mbb{R}_0$,
\bea
&&Y_s^{[1]} =y^{[1]}_1(s)X_s^{[1]}+y^{[1]}_0(s) \nn \\
&&Z_s^{[1]}=y^{[1]}_1(s)\sigma^{[0]}(s)  
\label{eq-hypo} \\
&&\psi_s^{[1]}(z)=y^{[1]}_1(s)\gamma^{[0]}(s,z)~. \nn
\eea
Here, $\Bigl(y^{[1]}_1(s), y^{[1]}_0(s), s\in[0,T]\Bigr)$ are the solutions to the following linear ODEs:
\bea
&&-\frac{dy^{[1]}_1(s)}{ds}=\bigl(\part_x b^{[0]}(s)+\part_y f^{[0]}(s)\bigr)y^{[1]}_1(s)+\part_x f^{[0]}(s), \nn \\
&&-\frac{dy^{[1]}_0(s)}{ds}=\part_y f^{[0]}(s)y^{[1]}_0(s)+\Bigl(\part_\ep b^{[0]}(s)+\part_z f^{[0]}(s)\sigma^{[0]}(s)+\part_u f^{[0]}(s)\Gamma^{[0]}(s)
\Bigr)y^{[1]}_1(s) \nn \\
\label{eq-L-ODE}
\eea
with the terminal conditions $y^{[1]}_1(T)=\part_x \xi^{[0]}$ and $y^{[1]}_0(T)=0$.
\begin{proof}
The existence of the unique solution for $\Theta^{[1]}$ is obvious from Lemmas \ref{ap-lemma-X-existence} and \ref{ap-lemma-Y-existence}.
Since the ODEs are linear with bounded coefficients as well the terminal conditions,  they obviously have 
bounded solutions $(y^{[1]}_0, y^{[1]}_1)$. 
The form of $Y^{[1]}$ is naturally expected from the linear structure of the BSDE and the order of $\ep$.
It automatically fixes the form of $Z^{[1]}$ and $\psi^{[1]}$. 
By applying It\^o-formula to the hypothesized $Y^{[1]}$ in (\ref{eq-hypo})
and using (\ref{eq-L-ODE}), one can directly confirm (\ref{eq-hypo}) gives the 
solution to the BSDE (\ref{eq-Y-1}). This also proves $\Theta^{[1]}\in \mbb{S}^p[0,T]^{\otimes 4}~\forall p\geq 2$.
Since the solution of the BSDE is unique, we are done.
\end{proof}
\end{lemma}

In the second order of $\ep$, we need to solve
\bea
&&X_s^{[2]}=\int_0^s \Bigl( \part_x b^{[0]}(r)X_r^{[2]}+\frac{1}{2}\part_x^2 b^{[0]}(r)(X_r^{[1]})^2
+\part_x\part_\ep b^{[0]}(r)X_r^{[1]}+\frac{1}{2}\part_\ep^2b^{[0]}(r)\Bigr)dr\nn \\
&&\quad+\int_0^s \part_x \sigma^{[0]}(r)X_r^{[1]}dW_r
+\int_{\mbb{R}_0}\part_x \gamma^{[0]}(r,z)X_r^{[1]}\wt{\mu}(dr,dz)
\label{eq-X-2}
\eea
and
\bea
&&Y_s^{[2]}=\part_x \xi^{[0]}X_T^{[2]}+\frac{1}{2}\part_x^2 \xi^{[0]}(X_T^{[1]})^2+\int_s^T \Bigl( \part_\Theta f^{[0]}(r)\Theta_r^{[2]}+\frac{1}{2}\part_\Theta^2 f^{[0]}(r)\Theta_r^{[1]}\Theta_r^{[1]}\Bigr)dr \nn \\
&&\qquad -\int_s^T Z_r^{[2]}dW_r-\int_s^t \psi_r^{[2]}(z)\wt{\mu}(dr,dz)~.
\label{eq-Y-2}
\eea
You can see that the dynamics of $X^{[2]}$ is linear in $X^{[2]}$ and contains $\{(X^{[1]})^j, j\leq 2\}$.
The BSDE for $\hat{\Theta}^{[2]}$ is linear in itself and contains $\{(\Theta^{[1]})^j,j\leq 2\}$.
Since we have seen $\hat{\Theta}^{[1]}$ is linear in $X^{[1]}$, 
the driver contains  $\{(X^{[1]})^j, j\leq 2\}$.
Suppose that $\hat{\Theta}^{[2]}$ is linear in $X^{[2]}$ and quadratic in $X^{[1]}$.
Then, one can check that this is also the case for  the driver of $Y^{[2]}$ and hence consistent with 
the initial assumption.
In fact, although it becomes a bit more tedious, 
one can prove the next lemma exactly in the same way as Lemma~\ref{lemma-ae-1}
by directly comparing the result of It\^o-formula with the driver of the BSDE.
\begin{lemma}
There exists a unique solution $\Theta^{[2]}$ to (\ref{eq-X-2}) and (\ref{eq-Y-2}) which 
belongs to $\mbb{S}^p[0,T]^{\otimes 4}~\forall p\geq 2$. $\hat{\Theta}^{[2]}$ is given by,
for $s\in[0,T]$ and $z\in\mbb{R}_0$,
\bea
&&Y_s^{[2]}=y^{[2]}_2(s)X_s^{[2]}+y^{[2]}_{1,1}(s)(X_s^{[1]})^2+y^{[2]}_1(s)X_s^{[1]}+y^{[2]}_0(s)\nn \\
&&Z_s^{[2]}=X^{[1]}_{s-}\Bigl(y^{[2]}_2(s)\part_x \sigma^{[0]}(s)+2y^{[2]}_{1,1}\sigma^{[0]}(s)\Bigr)
+y^{[2]}_1(s)\sigma^{[0]}(s)\nn \\
&&\psi_s^{[2]}(z)=X_{s-}^{[1]}\Bigl(y^{[2]}_2(s)\part_x \gamma^{[0]}(s,z)+2y^{[2]}_{1,1}(s)\gamma^{[0]}(s,z)\Bigr)
+y^{[2]}_{1,1}(s)(\gamma^{[0]}(s,z))^2+y^{[2]}_1(s)\gamma^{[0]}(s,z)~.\nn
\eea
Here, $\Bigl(y^{[2]}_2(s),y^{[2]}_{1,1}(s), y^{[2]}_1(s), y^{[2]}_0(s),s\in[0,T]\Bigr)$
are the solutions to the following linear ODEs:
\bea
-\frac{dy^{[2]}_2(s)}{ds}&=&\Bigl(\part_x b^{[0]}(s)+\part_y f^{[0]}(s)\Bigr)y^{[2]}_2(s)+\part_x f^{[0]}(s) \nn \\
-\frac{dy^{[2]}_{1,1}(s)}{ds}&=&\Bigl(2\part_x b^{[0]}(s)+\part_y f^{[0]}(s)\Bigr)y^{[2]}_{1,1}(s)
+\frac{1}{2}\part_x^2 f^{[0]}(s)\nn \\
&&+\frac{1}{2}\part_x^2 b^{[0]}(s)y^{[2]}_2(s)+\part_x\part_y f^{[0]}(s)y^{[1]}_1(s)+\frac{1}{2}\part_y^2 f^{[0]}(s)(y^{[1]}_1(s))^2 \nn 
\eea
\bea
-\frac{dy^{[2]}_{1}(s)}{ds}&=&\Bigl(\part_x b^{[0]}(s)+\part_y f^{[0]}(s)\Bigr)y^{[2]}_1(s)
+\part_x\part_\ep b^{[0]}(s)y^{[2]}_2(s)+2\part_\ep b^{[0]}(s)y^{[2]}_{1,1}(s)\nn \\
&&+\part_z f^{[0]}(s)\Bigl(y^{[2]}_2(s)\part_x \sigma^{[0]}(s)+2y^{[2]}_{1,1}(s)\sigma^{[0]}(s)\Bigr)\nn \\
&&+\part_u f^{[0]}(s)\Bigl(y^{[2]}_2(s)\part_x \Gamma^{[0]}(s)+2y^{[2]}_{1,1}(s)\Gamma^{[0]}(s)\Bigr)\nn \\
&&+\part_y^2 f^{[0]}(s)y^{[1]}_1(s)y^{[1]}_0(s)+\part_x\part_y f^{[0]}(s)y^{[1]}_0(s)\nn \\
&&+y^{[1]}_1(s)\Bigl(\part_x\part_zf^{[0]}(s)\sigma^{[0]}(s)+\part_x \part_u f^{[0]}(s)\Gamma^{[0]}(s)\Bigr)\nn \\
&&+(y^{[1]}_1(s))^2\Bigl(\part_y\part_z f^{[0]}(s)\sigma^{[0]}(s)+\part_y\part_u f^{[0]}(s)\Gamma^{[0]}(s)\Bigr)\nn \\
-\frac{dy^{[2]}_0(s)}{ds}&=&\part_y f^{[0]}(s)y^{[2]}_0(s)+
y^{[2]}_{1,1}(s)\Bigl((\sigma^{[0]}(s))^2+\int_{\mbb{R}_0}(\gamma^{[0]}(s,z))^2\nu(dz)\Bigr)\nn \\
&&+\frac{1}{2}\part_\ep^2 b^{[0]}(s)y^{[2]}_2(s)+\part_\ep b^{[0]}(s)y^{[2]}_1(s)
+y^{[2]}_1(s)\Bigl(\part_z f^{[0]}(s)\sigma^{[0]}(s)+\part_uf^{[0]}(s)\Gamma^{[0]}(s)\Bigr)\nn \\
&&+y^{[2]}_{1,1}(s)\part_u f^{[0]}(s)\int_{\mbb{R}_0}\rho(z)(\gamma^{[0]}(s,z))^2\nu(dz)
+\frac{1}{2}\part_y^2 f^{[0]}(s)(y^{[1]}_0(s))^2\nn \\
&&+(y^{[1]}_1(s))^2\Bigl(\frac{1}{2}\part_z^2 f^{[0]}(s)(\sigma^{[0]}(s))^2+\frac{1}{2}\part_u^2 f^{[0]}(s)(\Gamma^{[0]}(s))^2+\part_z\part_uf^{[0]}(s)\sigma^{[0]}(s)\Gamma^{[0]}(s)\Bigr)\nn \\
&&+(y^{[1]}_1(s)y^{[1]}_0(s))\Bigl(\part_y\part_z f^{[0]}(s)\sigma^{[0]}(s)+
\part_y\part_u f^{[0]}(s)\Gamma^{[0]}(s)\Bigr)\nn
\eea
with terminal conditions $y^{[2]}_2(T)=\part_x \xi^{[0]}$,
$y^{[2]}_{1,1}(T)=\frac{1}{2}\part_x^2 \xi^{[0]}$, $y^{[2]}_1(T)=y^{[2]}_0(T)=0$.
\end{lemma}

One can repeat the procedures to an any order $n\leq n_{\rm max}$. This can be checked in the following way.
By a simple modification of (\ref{eq-delxnY}) gives
\bea
Y^{[n]}_s=G_n+\int_s^T \Bigl\{ F_{n,r}+\part_\Theta f^{[0]}(r)\Theta^{[n]}_r\Bigr\}dr
-\int_s^T Z_r^{[n]}dW_r-\int_s^T\int_{\mbb{R}_0}\psi_r^{[n]}(z)\wt{\mu}(dr,dz) \nn
\eea
where
\bea
&&G_n:=\sum_{k=1}^n \sum_{\beta_1+\cdots+\beta_k=n,\beta_i\geq 1}\frac{1}{k!}\part_x^k \xi(X_T^{[0]})
\prod_{j=1}^k X_T^{[\beta_j]},\nn\\
&&F_{n,r}:=\sum_{k=2}^n \sum_{\beta_1+\cdots+\beta_k=n,\beta_i\geq 1}\sum_{i_x=0}^k
\sum_{i_y=0}^{k-i_x}\sum_{i_z=0}^{k-i_x-i_y}
\frac{\part_x^{i_x}\part_y^{i_y}\part_z^{i_z}\part_u^{k-i_x-i_y-i_z}f^{[0]}(r)}{i_x!
i_y!i_z!(k-i_x-i_y-i_z)!}\nn \\
&& \times \prod_{j_x=1}^{i_x}X_r^{[\beta_{j_x}]}
\prod_{j_y=i_x+1}^{i_x+i_y}Y_r^{[\beta_{j_y}]}
\prod_{j_z=i_x+i_y+1}^{i_x+i_y+i_z}Z_r^{[\beta_{j_z}]}
\prod_{j_u=i_x+i_y+i_z+1}^{k}
\int_{\mbb{R}_0} \rho(z)\psi^{[\beta_{j_u}]}_r(z)\nu(dz).\nn
\eea
From the shapes of $G_n, F_{n,r}$, one can confirm that $\hat{\Theta}^{[n]}_r$
is given by the polynomials 
\bea
\left\{\prod_{j=1}^k X_r^{[\beta_j]};
\beta_1+\cdots+\beta_k=m~(\beta_i\geq 1), ~k\leq m,~m\leq n
\right\} \nn
\eea
by induction. Since $\Theta^{[n]}$ appears only linearly both in the forward and backward SDEs
the relevant ODEs become always linear.

\subsection{A polynomial scheme}
\label{sec-polynomial}
We have just seen that the grading structure both for $\{X^{[n]}\}_{n \geq 0}$ and 
$\{\hat{\Theta}^{[n]}\}_{n\geq 0}$ played an important role. In particular, even if $\{\hat{\Theta}^{[n]}\}_{n\geq 0}$
has a grading structure, one cannot obtain the system of linear ODEs unless $\{X^{[n]}\}_{n\geq 0}$
shares the same features.
Suppose that the dynamics of $X^{t,x}$ is linear in itself. Then, one need not 
expand the forward SDE and thus one may obtain the expansion of $\hat{\Theta}^{t,x,\ep}$
in terms of polynomials of $X^{t,x}$. If this is the case, the ODEs for the associated coefficients
required in each order will be greatly simplified.

Let us consider the following forward-backward SDEs for $s\in[t,T]$:
\bea
\label{eq-X-poly}
&&X_s^{t,x}=x+\int_t^s\Bigl(b^0(r)+b^1(r)X_r^{t,x}\Bigr)dr+\int_t^s \Bigl(\sigma^0(r)+\sigma^1(r)X_r^{t,x}\Bigr)dW_r \nn \\
&&\quad +\int_s^t \int_E \Bigl(\gamma^0(r,z)+\gamma^1(r,z)X_{r-}^{t,x}\Bigr)  \wt{\mu}(dr,dz)  \\ 
&&Y_s^{t,x,\ep}=\xi(\ep X_T^{t,x})+\int_s^T f\Bigl(r, \ep X_r^{t,x},Y_r^{t,x,\ep}, Z_r^{t,x,\ep},
\int_{\mbb{R}_0}\rho(z)\psi_r^{t,x,\ep}(z)\nu(dz)\Bigr)dr\nn \\
&&\qquad-\int_s^T Z_r^{t,x,\ep}dW_r-\int_s^T \int_E \psi_r^{t,x,\ep}(z)\wt{\mu}(dr,dz)~.
\label{eq-Y-poly}
\eea
where $b^0:[0,T]\rightarrow \mbb{R}^d$, $b^1:[0,T]\rightarrow \mbb{R}^{d\times d}$,
$\sigma^0:[0,T]\rightarrow \mbb{R}^{d\times l}$, $\sigma^1:[0,T]\rightarrow \mbb{R}^{d\times d\times l}$,
$\gamma^0:[0,T]\times E \rightarrow \mbb{R}^{d\times k}$, $\gamma^1:[0,T]\times E \rightarrow \mbb{R}^{d\times d\times k}$
are measurable functions and $\xi,f$ are defined as before.

\begin{assumption}
\label{assumption-X-poly}
The functions $\{b^i(t),\sigma^i(t),\gamma^i(t,z)\}, i\in\{0,1\}$
are continuous. Furthermore, there exists some positive constant $K$ such that
$\Bigl(|b^i(t)|+|\sigma^i(t)|+|\gamma^i(t,z)|/\eta(z)\leq K\Bigr)$ for $i\in\{0,1\}$
uniformly in $(t,z)\in[0,T]\times E$.
\end{assumption}
With slight abuse of notation, let us use
$\Theta^{t,x,\ep}_r:=\Bigl(\ep X_r^{t,x}, Y_r^{t,x,\ep},Z_r^{t,x,\ep},\int_{\mbb{R}_0}\rho(z)\psi_r^{t,x,\ep}(z)\nu(dz)\Bigr)$
in this subsection. 

\begin{theorem}
Under Assumptions \ref{assumption-Y} and \ref{assumption-X-poly}, 
there exists a unique solution $\hat{\Theta}^{t,x,\ep}$ to the BSDE (\ref{eq-Y-poly}),
its classical derivative $\part_\ep^n \hat{\Theta}^{t,x,\ep}\in \calk^p[0,T]~\forall p\geq 2$ exists for every $0 \leq n\leq n^{\rm ae}$
and is given by the solution of the following BSDE:
\bea
&&\part_\ep^n Y_s^{t,x,\ep}=g_n (X_T^{t,x})^n+\int_s^T \Bigl\{h_{n,r}+\part_x^n f(r,\Theta^{t,x,\ep}_r)(X_r^{t,x})^n
+\part_{\hat{\Theta}}f(r,\Theta_r^{t,x,\ep})\part_\ep^n \hat{\Theta}_r^{t,x,\ep}\Bigr\}dr\nn \\
&&\qquad -\int_s^T \part^n_\ep Z_r^{t,x,\ep}dW_r-\int_s^T \int_E \part_\ep^n \psi^{t,x,\ep}_r(z)\wt{\mu}(dr,dz)\nn 
\eea
where $g_n:=\part_x^n \xi(\ep X_T^{t,x})$ and
\bea 
&&\hspace{-8mm}h_{n,r}:=n!\sum_{k=2}^n \sum_{i_x=0}^{k-1} 
\sum_{i_y=0}^{k-i_x}\sum_{i_z=0}^{k-i_x-i_y}\sum_{\beta_{i_x+1}+\cdots+\beta_k=n-i_x,~\beta_i\geq 1}
\frac{\part_x^{i_x}\part_y^{i_y}\part_z^{i_z}\part_u^{k-i_x-i_y-i_z}
f(r,\Theta_r^{t,x,\ep})}{i_x!i_y!i_z!(k-i_x-i_y-i_z)!} (X_r^{t,x})^{i_x}\nn \\
&&\hspace{-8mm}\times \prod_{j_y=i_x+1}^{i_x+i_y}\frac{1}{\beta_{j_y}!}\part_\ep^{\beta_{j_y}}Y^{t,x,\ep}_r
\prod_{j_z=i_x+i_y+1}^{i_x+i_y+i_z}\frac{1}{\beta_{j_z}!}\part_\ep^{\beta_{j_z}}Z^{t,x,\ep}_r
 \prod_{j_u=i_x+i_y+i_z+1}^k \frac{1}{\beta_{j_u}!}\int_{\mbb{R}_0}
\rho(z)\part_\ep^{\beta_{j_u}}\psi^{t,x,\ep}_r(z)\nu(dz)~.\nn
\eea
Moreover, for every $0\leq n\leq n_{\rm max}+1$, 
$\part_\ep^n \hat{\Theta}^{t,x,\ep}\in \mbb{S}^p[t,T]^{\otimes 3}~\forall p\geq2$.
The asymptotic expansion of $\hat{\Theta}^{t,x,\ep}$ with respect to $\ep$ satisfies,
with some positive constant $C_p$, that
\bea
\left|\left|\hat{\Theta}^{t,x,\ep}-\Bigl(\hat{\Theta}^{[0]}+\sum_{n=1}^N \ep^n \hat{\Theta}^{[n]}\Bigr)\right|\right|_{\mbb{S}^p[t,T]}
\leq  \ep^{N+1} C_p. \nn
\eea
for every $1\leq N\leq n_{\rm max}$.
\begin{proof}
One can follow the same arguments in Proposition~\ref{prop-delenY} and Theorem~\ref{theorem-AE}
by replacing $(X^{t,x,\ep})$ by $(\ep X^{t,x})$.
Since there is no $\ep$ dependence through $X^{t,x}$ in the expressions 
$Y_s^{t,x,\ep}=u(s,X_s^{t,x},\ep)$ and $Z_s^{t,x,\ep}=\part_x u(x,X_{s-}^{t,x},\ep)\sigma(s,X_{s-}^{t,x},\ep)$,
one-time differentiability with respect to $x$ and its polynomial growth property 
are enough to show recursively that $\part_\ep^n \hat{\Theta}^{t,x,\ep}\in \mbb{S}^{p}[t,T]$ for $\forall p\geq 2$.
\end{proof}
\end{theorem}

\begin{remark}
The above result also justifies the method proposed in Fujii (2015)~\cite{Fujii-poly} 
for the underlying $X$ with linear dynamics.
As for a general Affine-like process $X$ (such as $\sigma(x)=\sqrt{x}$), it is difficult to prove
within the current technique due to its non-Lipschitz nature.
\end{remark}

It is not difficult to see that $\bigl(\hat{\Theta}^{[n]}_s,s\in[t,T])$ 
is given by the unique solution to the following BSDE:
\bea
&&Y^{[n]}_s=\frac{1}{n!}\part_x^n \xi(0)(X_T^{t,x})^n+\int_s^T\Bigl\{\wt{h}_{n,r}+\frac{1}{n!}\part_x^n f^{[0]}(r)(X_r^{t,x})^n
+\part_{\hat{\Theta}}f^{[0]}(r)\hat{\Theta}^{[n]}_r\Bigr\}dr\nn \\
&&\qquad -\int_s^T Z_r^{[n]}dW_r-\int_s^T \int_E \psi_r^{[n]}(z)\wt{\mu}(dr,dz) 
\label{eq-ae-Y-poly}
\eea
where
\bea
&&\wt{h}_{n,r}:=\sum_{k=2}^n \sum_{i_x=0}^{k-1}\sum_{i_y=0}^{k-i_x}\sum_{i_z=0}^{k-i_x-i_y}
\sum_{\beta_{i_x+1}+\cdots+\beta_k=n-i_x, \beta_i\geq 1}
\frac{\part_x^{i_x}\part_y^{i_y}\part_z^{i_z}\part_u^{k-i_x-i_y-i_z}f^{[0]}(r)}
{i_x!i_y!i_z!(k-i_x-i_y-i_z)!}  (X_r^{t,x})^{i_x} \nn \\
&&\quad \times \prod_{j_y=i_x+1}^{i_x+i_y}Y_r^{[\beta_{j_y}]}
\prod_{j_z=i_x+i_y+1}^{i_x+i_y+i_z}Z_r^{[\beta_{j_z}]}
\prod_{j_u=i_x+i_y+i_z+1}^k \int_{\mbb{R}_0}\rho(z)\psi_r^{[\beta_{j_u}]}(z)\nu(dz)\nn
\eea
and $f^{[0]}(r):=f(r,0,Y^{[0]}_r,0,0)$.
Since $(i_x+\sum_{j_y}\beta_{j_y}+\sum_{j_z}\beta_{j_z}+\sum_{j_u}\beta_{j_u})=n$, one can recursively show 
that $\hat{\Theta}^{[n]}_r$ is given by the polynomials $\Bigl\{(X_r^{t,x})^j, 0\leq j\leq n\Bigr\}$ 
and every coefficient is determined by the system of linear ODEs as in Section~\ref{sec-implementation},
which we leave as a simple exercise.

\subsection*{An exponential L\'evy case}
In the reminder of this section, let us deal with a special example of 
an exponential (time-inhomogeneous) L\'evy dynamics for $X$.
Let us put $m=d=l=k=1$ and $t=0$ for simplicity and consider $b^0=\sigma^0=\gamma^0=0$
\bea
\label{eq-X-exp}
X_s=x+\int_t^s X_r\Bigl(b(r)dr+\sigma(r)dW_r\Bigr)+\int_s^t \int_{\mbb{R}_0}X_{r-}\gamma(r,z)\wt{\mu}(dr,dz)
\eea
with $b:=b^1, \sigma:=\sigma^1,\gamma:=\gamma^1$ in (\ref{eq-X-poly}). We omit the superscript denoting the initial data $(0,x)$.
Let us introduce the notations: $q(s,j):=\int_{\mbb{R}_0}(\gamma(s,z))^j \nu(dz)$ for $j\geq 2$,
$\Gamma(s,j):=\int_{\mbb{R}_0}\rho(z)\bigl[(1+\gamma(s,z))^j-1]\nu(dz)$ for $j\geq 1$
and $C_{n,j}:=n!/(j!(n-j)!)$ for $j\leq n, n\geq 2$.
\begin{theorem}
Under Assumptions \ref{assumption-X}, \ref{assumption-X-poly}, $m=d=l=k=1$ and $t=0$,
the asymptotic expansion of the forward-backward SDEs (\ref{eq-X-exp}) and (\ref{eq-Y-poly})
is given by, for $s\in[0,T]$, 
\bea
\label{eq-y-0}
&&Y^{[0]}_s=\xi(0)+\int_s^T f(r,0,Y^{[0]}_r,0,0) dr \\
&&Z^{[0]}=\psi^{[0]}=0\nn
\eea
and, for $1\leq n\leq n_{\rm max}$, 
\bea
&&Y_s^{[n]}=(X_s)^ny^{[n]}(s)\nn \\
&&Z_s^{[n]}=(X_{s-})^n y^{[n]}(s)n\sigma(s)\nn \\
&&\psi_s^{[n]}(z)=(X_{s-})^ny^{[n]}(s)\bigl[(1+\gamma(s,z))^n-1\bigr] \nn 
\eea
where the  functions $\{y^{[j]}(s),s\in[0,T]\}_{1\leq j\leq n}$
are determined recursively by the following system of linear ODEs:
\bea
-\frac{dy^{[n]}(s)}{ds}&=&\Bigl( nb(s)+\frac{1}{2}n(n-1)\sigma^2(s)+\sum_{j=2}^n C_{n,j}q(s;j)+\part_y f^{[0]}(s)\nn \\
&&+\part_z f^{[0]}(s)n\sigma(s)+\part_u f^{[0]}(s)\Gamma(s;n)\Bigr)y^{[n]}(s)+\frac{1}{n!}\part_x^n f^{[0]}(s)\nn\\
&&+\sum_{k=2}^n\sum_{i_x=0}^{k-1}\sum_{i_y=0}^{k-i_x-i_y}
\sum_{\beta_{i_x+1}+\cdots+\beta_k=n-i_x,\beta_i\geq 1}\left\{ 
\frac{\part_x^{i_x}\part_y^{i_y}\part_z^{i_z}\part_u^{k-i_x-i_y-i_z}f^{[0]}(s)}{i_x!i_y!i_z!(k-i_x-i_y-i_z)!} \right. \nn\\
&& \times \prod_{j_y=i_x+1}^{i_x+i_y}\Bigl(y^{[\beta_{j_y}]}(s)\Bigr)
\prod_{j_z=i_x+i_y+1}^{i_x+i_y+i_z}\Bigl(\beta_{j_z}\sigma(s)y^{[\beta_{j_z}]}(s)\Bigr) \nn\\ 
&&\times \left. \prod_{j_u=i_x+i_y+i_z+1}^k\Bigl(\Gamma(s;\beta_{j_u})y^{[\beta_{j_u}]}(s)\Bigr)\right\} \nn
\eea
with a terminal condition $y^{[n]}(T)=\part_x^n \xi(0)/n!$ for every $n$.
Here, $f^{[0]}(r)$ is defined by $f(r,0,Y^{[0]}_r,0,0)$ using $Y^{[0]}$ determined by (\ref{eq-y-0}).
\begin{proof}
If one supposes the form of the solution as $Y^{[n]}_s=(X_s)^n y^{[n]}(s)$, then $Z^{[n]}$ and $\psi^{[n]}$
must have the form as given. Comparing the result of It\^o formula applied to $X^n y^{[n]}$
and the form of the BSDE (\ref{eq-ae-Y-poly}) substituted by the hypothesized form of $\{\hat{\Theta}^{[\beta]}\}_{\beta\leq n}$,
one obtains the system of ODEs given above.
Since every ODE is linear, there exists a solution for every $y^{[n]}$, $1\leq n\leq n_{\rm max}$.
Since the solution of the BSDE is unique, this must be the desired solution.
\end{proof}
\end{theorem}

\subsubsection*{Remark}
It is interesting to observe the difference from the linearization scheme proposed in \cite{FT-analytic} for a Brownian setup.
There, the BSDE is expanded around a linear driver in the first step. 
Then in the second step 
the resultant set of linear BSDEs are evaluated by the small-variance asymptotic expansion of the forward SDE, or by
the interacting particle simulation method proposed in Fujii \& Takahashi (2015)~\cite{FT-particle}.
Hence,  in order for the scheme of \cite{FT-analytic} works well, it requires the smallness of the non-linear terms in the driver $f$,
although it naturally arises in many  applications.
Furthermore, due to the presence of large number of conditional expectations, calculating them
analytically without invoking the particle simulation technique~\cite{FT-particle} is unrealistic in most of the 
practical situations.

On the other hand, in the current scheme, the expansion of the driver is not directly performed 
and the significant part of non-linearity is taken into account 
at the zero-th order around the mean dynamics of the forward SDE as observed in (\ref{eq-Y-0th}).
The effects of the stochasticity from the forward SDE are then taken into account 
perturbatively around this ``mean" solution. 
Therefore, the current scheme is expected to be more advantageous when 
there exists significant non-linearity in the driver.
Furthermore, the special grading structure of approximating FBSDEs makes them explicitly solvable 
by ODEs without using any Monte-Carlo simulation. Since the approximate solution of $(Y,Z,\psi(\cdot))$ 
is explicitly given as a polynomial
in the stochastic flows of $X$, one can obtain not only the current value $(Y_0,Z_0,\psi_0(\cdot))$
but also its evolution by simply simulating the flows of $X$ (or $X$ itself for the polynomial case).
Some numerical examples and empirical error estimates are available in Fujii (2015)~\cite{Fujii-poly}
based on this property for a certain class of models.

\appendix
\section{Useful a priori estimates: forward SDEs}
Let us summarize the useful a priori estimates for FSDEs with jumps.
The following result taken from Lemma 5-1 of Bichteler, Gravereaux and Jacod (1987)~\cite{Bichteler}
is essential for analysis of a $\sigma$-finite random measure.
\begin{lemma}
\label{ap-lemma-BGJ}
Let $\eta:\mbb{R}\rightarrow \mbb{R}$ be defined by $\eta(z)=1\wedge |z|$.
Then, for $\forall p\geq 2$, there exists a constant $\del_p$ depending on $p,T, m,k$ 
such that
\bea
\label{BDG-nu}
\mbb{E}\left[\sup_{t\in[0,T]}\Bigl|\int_0^t\int_E U(s,z)\wt{\mu}(ds,dz)\Bigr|^p\right]
\leq \del_p \int_0^T \mbb{E}|L_s|^p ds
\eea 
if $U$ is an $\mbb{R}^{m\times k}$-valued $\calp\otimes \cale$-measurable function on $\Omega\times[0,T]\times E$
and $L$ is a predictable process satisfying $|U_{\cdot,i}(\omega,s,z)|\leq L_s(\omega)\eta(z)$ for
each column $1\leq i\leq k$. 
\end{lemma}
Since $\int_E \eta(z)^p\nu(dz)<\infty$ for $\forall p\geq 2$, the above lemma tells that 
one can use a BDG-like inequality with a compensator $\nu$ whenever the integrand of the 
random measure divided by $\eta$ is dominated by some integrable random variable.
The following result from 
Chapter 1 Section 9 Lemma 6 of Liptser \& Shiryayev (1989)~\cite{Shiryayev} or
Lemma 2.1 of Dzhaparidze \& Valkeila (1990)~\cite{Dzhaparidze} is also important.
\begin{lemma}
\label{ap-lemma-nu-vs-mu}
Let $\psi$ belong to $\mbb{H}^2_{\nu}[0,T]$. Then, for $p\geq 2$, 
there exists some constant $C_p>0$ depending only on $p$ such that
\bea
\mbb{E}\Bigl(\int_0^T \int_E |\psi_s(z)|^2\nu(dz)ds\Bigr)^{p/2} \leq
C_p \mbb{E}\Bigl(\int_0^T \int_E |\psi_s(z)|^2\mu(ds,dz)\Bigr)^{p/2}~.\nn
\eea
\end{lemma}
\vspace{5mm}

For $t_1\leq t_2\leq T$ and $\mbb{R}^d$-valued $\calf_{t_i}$-measurable random variable $x^i$,
let us consider $\{X^i_t,t\in[t_i,T]\}_{1\leq i\leq 2}$ as a solution of the following SDE:
\bea
\label{ap-eq-X}
X_t^i=x^i+\int_{t_i}^t\wt{b}^i(s,X_s^i)ds+\int_{t_i}^t \wt{\sigma}^i(s,X_s^i)dW_s
+\int_{t_i}^t\int_E \wt{\gamma}^i(s,X_{s-}^i,z)\wt{\mu}(ds,dz)
\eea
where $\wt{b}^i: \Omega\times[0,T]\times\mbb{R}^d\rightarrow \mbb{R}^d$,
$\wt{\sigma}^{i}: \Omega \times[0,T]\times \mbb{R}^d \rightarrow \mbb{R}^{d\times l}$,
and $\wt{\gamma}^i: \Omega\times [0,T]\times \mbb{R}^d\times E\rightarrow \mbb{R}^{d\times k}$.

\begin{assumption}
\label{ap-assumption}
For $i\in\{1,2\}$, the map $(\omega,t) \mapsto \wt{b}^i(\omega, t,\cdot)$ is $\mbb{F}$-progressively measurable,
$(\omega,t) \mapsto \wt{\sigma}^i(\omega, t,\cdot), \wt{\gamma}^i(\omega,t,\cdot)$ are $\mbb{F}$-predictable, and there exists
some constant $K>0$ such that, for every $x,x^\prime \in \mbb{R}^d$ and $z\in E$,
\bea
&&|\wt{b}^i(\omega,t,x)-\wt{b}^i(\omega,t,x^\prime)|+|\wt{\sigma}^i(\omega,t,x)-\wt{\sigma}^i(\omega,t,x^\prime)|
\leq K |x-x^\prime| \nn \\
&&|\wt{\gamma}_{\cdot,j}^i(\omega,t,x,z)-\wt{\gamma}_{\cdot,j}^i(\omega,t,x^\prime,z)|\leq K\eta(z) |x-x^\prime|,
\quad 1\leq j\leq k\nn
\eea
$d\mbb{P}\otimes dt$-a.e. in $\Omega\times[0,T]$.
Furthermore, for some $p\geq 2$, 
\bea
\mbb{E}\left[|x^i|^p+\Bigl(\int_{t_i}^T |\wt{b}^i(s,0)|ds\bigr)^p+\Bigl(\int_{t_i}^T |\wt{\sigma}^i(s,0)|^2ds\Bigr)^{p/2}
+\int_{t_i}^T |L^i_s|^pds\right]<\infty \nn
\eea
where $L^i$ is some $\mbb{F}$-predictable process satisfying $|\wt{\gamma}^i_{\cdot,j}(\omega,t,0,z)|\leq L^i_t(\omega)\eta(z)$
for every column vector $\{\wt{\gamma}^i_{\cdot,j},~1\leq j\leq k\}$.
\end{assumption}
The following lemma is an extension of Lemma A.1 given in \cite{Bouchard-Elie}
to a $\sigma$-finite measure by using (\ref{BDG-nu}).
\begin{lemma}
\label{ap-lemma-X-existence}
Under Assumption~\ref{ap-assumption}, the SDE (\ref{ap-eq-X}) has a unique solution and there exists
some constant $C_p>0$ such that, 
\bea
&&||X^i||^p_{\mbb{S}^p_d[t_i,T]}\leq C_p\mbb{E}\left[|x^i|^p+
\Bigl(\int_{t_i}^T|\wt{b}^i(s,0)|ds\Bigr)^p \right. \nn \\
&&\hspace{20mm}\left.+\Bigl(\int_{t_i}^T |\wt{\sigma}^i(s,0)|^2ds\Bigr)^{p/2}+
\int_{t_i}^T |L^i_s|^p ds\right] 
\label{ap-X-supnorm}
\eea
and, for all $t_i\leq s\leq t\leq T$, 
\be
\mbb{E}\left[\sup_{s\leq u\leq t}|X_u^i-X_s^i|^p\right]\leq C_p A_p^i|t-s| 
\label{ap-X-tspread}
\ee
where 
\be
A_p^i:=\mbb{E}\left[|x^i|^p+||\wt{b}^i(\cdot,0)||_{[t_i,T]}^p
+||\wt{\sigma}^i(\cdot,0)||_{[t_i,T]}^p+||L^i||^p_{[t_i,T]}\right]~.\nn
\ee
Moreover, for $t_2\leq t\leq T$,
\bea
&&||\del X||^p_{\mbb{S}^p_d[t_2,T]}\leq C_p \Bigl(\mbb{E}|x^1-x^2|^p+A_p^1|t_2-t_1|\Bigr)\nn \\
&&\qquad+C_p\mbb{E}\left[ \Bigl(\int_{t_2}^T |\del \wt{b}_t|dt\Bigr)^p+
\Bigl(\int_{t_2}^T |\del\wt{\sigma}_t|^2dt\Bigr)^{p/2}+
\int_{t_2}^T |\del L_t|^pdt\right] 
\label{ap-X-stability}
\eea
where $\del X:=X^1-X^2$, $\del\wt{b}_\cdot:=(\wt{b}^1-\wt{b}^2)(\cdot,X^1_\cdot)$,
$\del\wt{\sigma}_\cdot:=(\wt{\sigma}^1-\wt{\sigma}^2)(\cdot,X^1_\cdot)$
and $\del L$ is a predictable process satisfying
$|\del\wt{\gamma}|(\omega, t,z)\leq \del L_t(\omega)~\eta(z)$,  $d\mbb{P}\otimes dt$-a.e. in $\Omega\times[0,T]$,
where $\del\wt{\gamma}(\omega,t,z):=(\wt{\gamma}^1-\wt{\gamma}^2)(\omega,t,X_{t-}^1(\omega),z)$.
\begin{proof}
The existence of a unique solution is given in pp.237 of Gikhman \& Skorohod (1972)~\cite{Gikhman} or Section 6.2 of Applebaum (2009)~\cite{Applebaum}, for example.
For the sake of completeness, let us give a sketch of proof for the other estimates.
 
Set a sequence of stopping times $\Bigl(\tau_n:=\inf\{t\geq t_i; |X_s^i|\geq n\}\wedge T,~n\in\mbb{N}\Bigr)$. Then,
using the fact that $|\wt{\gamma}^i(s,X_{s-}^i,z)|\leq (L^i_s+K|X_{s-}^i|)\eta(z)$, Lemma~\ref{ap-lemma-BGJ} and 
the Burkholder-Davis-Gundy (BDG) inequality,
one obtains
\bea
\mbb{E}|X_{\tau_n}^i|^p&\leq& C_p\int_{t_i}^{\tau_n}\mbb{E}|X_s^i|^p ds \nn \\
&+& C_p\mbb{E}\left[|x^i|^p+
\Bigl(\int_{t_i}^{\tau_n}|\wt{b}^i(s,0)|ds\Bigr)^p+\Bigl(\int_{t_i}^{\tau_n}|\wt{\sigma}^i(s,0)|^2ds\Bigr)^{p/2}
+\int_{t_i}^{\tau_n}|L_s^i|^p ds\right]~.\nn 
\eea
Using the Gronwall inequality and passing to the limit $n\rightarrow \infty$, one obtains
the estimate for $\Bigl(\sup_{t\in[t_i,T]}\mbb{E}|X^i_t|^p\Bigr)$. Using the BDG inequality and Lemma~\ref{ap-lemma-BGJ} once again,
one obtains the first estimate (\ref{ap-X-supnorm}). 
A similar analysis yields
\bea
&&\mbb{E}\sup_{u\in[s,t]}|X_u^i-X_s^i|^p\leq C_p\mbb{E}\left[\Bigl(\int_s^t|\wt{b}^i(r,0)|dr\Bigr)^p
+\Bigl(\int_s^t |\wt{\sigma}^i(r,0)|^2 dr\Bigr)^{p/2}+\int_s^t |L^i_r|^p dr\right]\nn \\
&&\qquad\qquad+C_p(t-s)\mbb{E}||X^i||^p_{[t_i,T]},\nn
\eea
which gives second estimate (\ref{ap-X-tspread}).

As for the last estimate (\ref{ap-X-stability}), notice first that
\be
|\wt{\gamma}^1-\wt{\gamma}^2|(s,X_{s-}^1,z)\leq (L_s^1+L_s^2+2K|X_{s-}^1|)\eta(z)~\nn.
\ee
Since $X^1\in\mbb{S}^p$, there exists a predictable process $\del L$ satisfying
$|\wt{\gamma}^1-\wt{\gamma}^2|(s,X_{s-}^1,z)\leq \del L_s \eta(z)$, $d\mbb{P}\otimes ds$-a.e. 
and $\int_{t_2}^T \mbb{E}|\del L_r|^p dr<\infty$ as desired.
Separating the integration range, applying the BDG inequality and Lemma~\ref{ap-lemma-BGJ}, one obtains
\bea
\mbb{E}||\del X||^p_{[t_2,t]}&\leq& C_p\mbb{E}\left[|x^1-x^2|^p+
\Bigl(\int_{t_1}^{t_2}|\wt{b}^1(s,0)|ds\Bigr)^p+\Bigl(\int_{t_1}^{t_2}|\wt{\sigma}^1(s,0)|^2 ds\Bigr)^{p/2}\right. \nn \\
&&\left.+\int_{t_1}^{t_2}|L_s^1|^pds+(t_2-t_1)||X^1||^p_{[t_1,t_2]}\right]
+C_p\mbb{E}\left[\int_{t_2}^t |\del X_s|^p ds \right.\nn\\
&&\left. +\Bigl(\int_{t_2}^t |\del \wt{b}_s|ds\Bigr)^p+\Bigl(\int_{t_2}^t |\del \wt{\sigma}_s|^2 ds\Bigr)^{p/2}
+\int_{t_2}^t |\del L_s|^p ds\right]~.\nn
\eea
Using the first two results and the Gronwall inequality, one obtains (\ref{ap-X-stability}).
\end{proof}
\end{lemma}

\subsubsection*{Remark}
Note that when $p=2$, one can replace $\int^\cdot |L^i_s|^2 ds$ (resp. $\int^\cdot |\del L_s|^2 ds$)
by $\int^\cdot \int_E |\wt{\gamma}^i(s,0,z)|^2\nu(dz)ds$ (resp. $\int^\cdot \int_E |\del \wt{\gamma}(s,z)|^2
\nu(dz)ds$) by simply applying the BDG inequality.
Furthermore, when the compensator is finite $\nu(E)<\infty$,
the above replacement is possible for any $\forall p\geq 2$
thanks to Lemma~\ref{ap-lemma-finite} (see below).
\\

\section{Useful a priori estimates: BSDEs}
Consider the following BSDE:
\bea
Y_t=\wt{\xi}+\int_t^T \wt{f}(s,Y_s,Z_s,\psi_s)ds-\int_t^T Z_s dW_s-\int_t^T \int_E \psi_s(z)\wt{\mu}(ds,dz)~,
\label{eq-Y-apriori}
\eea
where $\xi:\Omega\rightarrow \mbb{R}^m, \wt{f}:\Omega\times [0,T]\times \mbb{R}^m\times \mbb{R}^{m\times l}\times 
\mbb{L}^2(E,\cale,\nu;\mbb{R}^m)\rightarrow \mbb{R}^m$.
In this section, we use $\langle \cdot, \cdot\rangle$ to denote an inner product of $m$-dimensional vectors for clarity.
\begin{assumption}
\label{assumption-monotone-growth}
(i) $\wt{\xi}$ is $\calf_T$-measurable  and the map $(\omega,t)\mapsto \wt{f}(\omega,t,\cdot)$
is $\mbb{F}$-progressively measurable. There exists a solution $(Y,Z,\psi)$ to the BSDE (\ref{eq-Y-apriori}).\\
(ii) For $\forall \lambda\in(0,1)$, there exist an $\mbb{F}$-progressively measurable continuous process with 
bounded variation $(V^{\lambda}_s, s\in[0,T])$ with $V^\lambda_0=0$
and an $\mbb{F}$-progressively measurable increasing process $(N_s^\lambda,s\in[0,T])$ with $N_0=0$ 
such that, as a signed measure on $\mbb{R}_+$,
\bea 
\langle Y_s,\wt{f}(s,Y_s,Z_s,\psi_s)\rangle ds\leq |Y_s|^2 dV_s^{\lambda}+|Y_s|dN^\lambda_s+\lambda(|Z_s|^2+||\psi_s||^2_{\mbb{L}^2(E)})ds~. \nn
\eea
(iv) There exists some $p\geq 2$ such that
$\mbb{E}\left[\bigl|\bigl|e^{V^{\lambda}}Y\bigr|\bigr|^p_T+\Bigl(\int_0^T e^{V_s^\lambda} dN_s^\lambda \Bigr)^p\right]<\infty~$
is satisfied  for every $\forall \lambda\in(0,1)$.
\end{assumption}

\begin{lemma}
\label{lemma-monotone-apriori}
Suppose Assumption~\ref{assumption-monotone-growth} hold true. Then, there exists some $\exists \lambda\in(0,1)$
such that the following inequality is satisfied;
\bea
&&\mbb{E}||e^{V^{\lambda}}Y||^p_T+\mbb{E}\left(\int_0^T e^{2V_s^{\lambda}}|Z_s|^2 ds\right)^{\frac{p}{2}}+\mbb{E}\left(\int_0^T\int_E 
e^{2V_s^\lambda}|\psi_s(z)|^2\mu(ds,dz)\right)^{
\frac{p}{2}}\nn \\
&&\quad+\mbb{E}\left(\int_0^T\int_E e^{2V_s^\lambda} |\psi_s(z)|^2\nu(dz)ds\right)^{\frac{p}{2}}
\leq C_{p,\lambda} \mbb{E}\left[e^{pV^\lambda_T} |\wt{\xi}|^p+\Bigl(\int_0^T e^{V_s^\lambda}dN_s^\lambda \Bigr)^p\right]~,\nn
\eea
where $C_{p,\lambda}$ is a positive constant depending only on $p, \lambda$.
\begin{proof}
The following proof is an improvement of Proposition~2 of Kruse \& Popier (2015)~\cite{Kruse}
by following the idea of Proposition 6.80 of Pardoux \& Rascanu (2014)~\cite{Pardoux-Rascanu},
which yields a slightly sharper a priori estimate for $p\geq 2$.
\\\\
{\it First step:}
Introduce a sequence of stopping times with $n\in\mbb{N}$,
\bea
\tau_n&:=&\inf\Bigl\{t\geq 0;\int_0^te^{2V_s^\lambda} |Z_s|^2 ds+\int_0^t\int_E e^{2V_s^\lambda} |\psi_s(z)|^2
\bigl(\mu(ds,dz)+\nu(dz)ds\bigr) \nn \\
&&\quad+||e^{V^\lambda}Y||_t+\int_0^te^{V_s^\lambda} dN_s^\lambda
\geq n\Bigr\}\wedge T. \nn
\eea
One obtains by applying It\^o formula
\bea
&&|Y_0|^2+\int_0^{\tau_n} e^{2V_s^\lambda}|Z_s|^2 ds+\int_0^{\tau_n}\int_E e^{2V_s^\lambda}|\psi_s(z)|^2\mu(ds,dz)\nn \\
&&=e^{2V_{\tau_n}^\lambda}|Y_{\tau_n}|^2+\int_0^{\tau_n}e^{2V_s^\lambda} 2\Bigl(\langle Y_s,
\wt{f}(s,Y_s,Z_s,\psi_s)\rangle ds-|Y_s|^2 dV_s^\lambda\Bigr)\nn \\
&&\quad -\int_0^{\tau_n}e^{2V_s^\lambda}2 \langle Y_s,Z_s dW_s\rangle-\int_0^{\tau_n}e^{2V_s^\lambda}
2\langle Y_{s-},\psi_s(z)\rangle \wt{\mu}(ds,dz)~\nn \\
&&\leq e^{2V_{\tau_n}^\lambda}|Y_{\tau_n}|^2+\int_0^{\tau_n}e^{2V_s^\lambda} 2\Bigl(|Y_s|dN_s^\lambda
+\lambda (|Z_s|^2+||\psi_s||^2_{\mbb{L}^2(E)})ds\Bigr)\nn \\
&&\quad -\int_0^{\tau_n}e^{2V_s^\lambda}2 \langle Y_s,Z_s dW_s\rangle-\int_0^{\tau_n}\int_E e^{2V_s^\lambda}
2\langle Y_{s-},\psi_s(z)\rangle \wt{\mu}(ds,dz)~\nn.
\eea

The BDG (or Davis when $p=2$) inequality yields, with some positive constant $C_p$
depending only on $p$,
\bea
&&\mbb{E}\left[\Bigl(\int_0^{\tau_n}e^{2V_s^\lambda}|Z_s|^2 ds\Bigr)^{\frac{p}{2}}+
\Bigl(\int_0^{\tau_n}\int_E  e^{2V_s^{\lambda}}|\psi_s(z)|^2\mu(ds,dz)\Bigr)^{\frac{p}{2}}\right]\nn \\
&&\leq C_p\mbb{E}\left[ ||e^{V^\lambda} Y||^p_{\tau_n}+\Bigl(\int_0^{\tau_n}e^{V_s^\lambda}dN_s^\lambda\Bigr)^p\right]\nn \\
&&\quad +\lambda^{\frac{p}{2}}C_p\mbb{E}\left[\Bigl(\int_0^{\tau_n}e^{2V_s^\lambda}|Z_s|^2 ds\Bigr)^{\frac{p}{2}}+
\Bigl(\int_0^{\tau_n} e^{2V_s^\lambda}||\psi_s||_{\mbb{L}^2(E)}ds\Bigr)^{\frac{p}{2}}\right]\nn \\
&&\quad+C_p\mbb{E}\left[\Bigl(\int_0^{\tau_n}e^{4V_s^\lambda}|Y_s|^2|Z_s|^2 ds\Bigr)^{\frac{p}{4}}
+\Bigl(\int_0^{\tau_n}\int_Ee^{4V_s^\lambda}|Y_s|^2|\psi_s(z)|^2\mu(ds,dz)\Bigr)^{\frac{p}{4}}\right]~.\nn
\eea
With an arbitrary constant $\ep>0$, one has
\bea
&&C_p\mbb{E}\left[\Bigl(\int_0^{\tau_n}e^{4V_s^\lambda}|Y_s|^2|Z_s|^2 ds\Bigr)^{\frac{p}{4}}\right] 
\leq C_p\mbb{E}\left[||e^{V^\lambda}Y||^{\frac{p}{2}}_{\tau_n}\Bigl(\int_0^{\tau_n}e^{2V_s^\lambda}|Z_s|^2ds\Bigr)^{\frac{p}{4}}
\right]\nn \\
&&\qquad \leq \frac{C_p^2}{4\ep}\mbb{E}\Bigl[||e^{V^\lambda}Y||^p_{\tau_n}\Bigr]+\ep\mbb{E}\left[\Bigl(\int_0^{\tau_n} e^{2V_s^\lambda}|Z_s|^2 ds\Bigr)^{\frac{p}{2}}\right]\nn 
\eea
and similarly
\bea
&&C_p\mbb{E}\left[\Bigl(\int_0^{\tau_n}\int_E e^{4V_s^\lambda}|Y_s|^2|\psi_s(z)|^2 \mu(ds,dz)\Bigr)^{\frac{p}{4}}\right] \nn \\
&&\qquad \leq \frac{C_p^2}{4\ep}\mbb{E}\Bigl[||e^{V^\lambda}Y||^p_{\tau_n}\Bigr]+\ep\mbb{E}\left[\Bigl(\int_0^{\tau_n}\int_E e^{2V_s^\lambda}|\psi_s(z)|^2 \mu(ds,dz)\Bigr)^{\frac{p}{2}}\right].\nn
\eea

Thus, one obtains
\bea
&&(1-\ep-\lambda^{\frac{p}{2}} C_p)\mbb{E}\Bigl(\int_0^{\tau_n}e^{2V_s^\lambda}|Z_s|^2 ds\Bigr)^{\frac{p}{2}}
+(1-\ep)\mbb{E}\Bigl(\int_0^{\tau_n}\int_E e^{2V_s^\lambda}|\psi_s(z)|^2\mu(ds,dz)\Bigr)^{\frac{p}{2}}\nn \\
&&-\lambda^{\frac{p}{2}} C_p\mbb{E}\Bigl(\int_0^{\tau_n}\int_E e^{2V_s^\lambda}|\psi_s(z)|^2\nu(dz)ds\Bigr)^{\frac{p}{2}}
\leq C_p^\prime \mbb{E}\left[ ||e^{V^\lambda} Y||^p_{\tau_n}+\Bigl(\int_0^{\tau_n}e^{V_s^\lambda}dN_s^\lambda\Bigr)^p\right]~.\nn
\eea
Firstly, choose some $\ep\in(0,1)$. Then, by Lemma~\ref{ap-lemma-nu-vs-mu}, there exists a $\lambda\in(0,1)$ depending only on $p$ 
so that the 3rd term  is absolutely smaller than the 2nd term.
Redefining the coefficients and  passing to the limit $\tau_n\rightarrow T$ yields
\bea
&&\hspace{-7mm}\mbb{E}\left[\Bigl(\int_0^{T} e^{2V_s^\lambda}|Z_s|^2 ds\Bigr)^{\frac{p}{2}}+
\Bigl(\int_0^{T}\int_E e^{2V_s^\lambda}|\psi_s(z)|^2\mu(ds,dz)\Bigr)^{\frac{p}{2}}\right]\nn \\
&&\hspace{-7mm}+\mbb{E}\left[\Bigl(\int_0^{T}\int_E e^{2V_s^\lambda}|\psi_s(z)|^2\nu(dz)ds\Bigr)^{\frac{p}{2}}\right] \leq C_{p,\lambda}\mbb{E}\left[||e^{V^{\lambda}}Y||^p_T+\Bigl(\int_0^T e^{V_s^\lambda}dN_s^\lambda\Bigr)^p\right].
\label{eq-first-result}
\eea
\\
{\it Second step:} 
Put $\theta(y):=|y|^p$. Then, It\^o formula yields
\bea
&&d(e^{pV_s^\lambda }|Y_s|^p)=e^{pV_s^\lambda }\Bigl(p|Y_s|^pdV_s^\lambda+p|Y_{s-}|^{p-2}\langle Y_{s-},dY_s\rangle+
\frac{1}{2}{\rm Tr}(\part_y^2 \theta(Y_s)Z_sZ_s^\top)ds\Bigr)\nn \\
&&+\int_E e^{pV_s^\lambda}\Bigl(|Y_{s-}+\psi_s(z)|^p-|Y_{s-}|^p-p|Y_{s-}|^{p-2}\langle Y_{s-},\psi_s(z)\rangle \Bigr)\mu(ds,dz).\nn
\eea
Using the same sequence of stopping times $(\tau_n)_{n\in\mbb{N}}$,
\bea
&&e^{pV_t^\lambda }|Y_t|^p=e^{pV_{\tau_n}^\lambda}|Y_{\tau_n}|^p+\int_t^{\tau_n}e^{pV_s^\lambda}p|Y_s|^{p-2}
\Bigl(\langle Y_s,\wt{f}(s,Y_s,Z_s,\psi_s)\rangle ds-|Y_s|^2 dV_s^\lambda \Bigr)\nn \\
&&-\int_t^{\tau_n}e^{pV_s^\lambda }\frac{1}{2}{\rm Tr}(\part_y^2\theta(Y_s)Z_sZ_s^\top)ds\nn \\
&&-\int_t^{\tau_n}\int_E e^{pV_s^\lambda }\Bigl(|Y_{s-}+\psi_s(z)|^p-|Y_{s-}|^p-p|Y_{s-}|^{p-2}\langle Y_{s-},\psi_s(z)\rangle\Bigr)\mu(ds,dz)\nn \\
&&-\int_t^{\tau_n}e^{pV_s^\lambda }p|Y_s|^{p-2}\langle Y_s,Z_s dW_s\rangle-\int_t^{\tau_n}\int_E
e^{pV_s^\lambda }p|Y_{s-}|^{p-2}\langle Y_{s-},\psi_s(z)\rangle\wt{\mu}(ds,dz)~.\nn
\eea
Let us mention the fact that
\bea
&&{\rm Tr}(\part_y^2 \theta(Y_s)Z_sZ_s^\top)\geq p|Y_s|^{p-2}|Z_s|^2,\nn \\
&&|Y_{s-}+\psi_s^i(z)|^p-|Y_{s-}|^p-p|Y_{s-}|^{p-2}\langle Y_{s-},\psi_s^i(z)\rangle
\geq p(p-1)3^{1-p}|Y_{s-}|^{p-2}|\psi_s^i(z)|^2,\nn
\eea
for every $i\in\{1,\cdots,k\}$. The latter is obtained by evaluating the residual of Taylor formula~\cite{Kruse}.
Setting $\kappa_p:=\min\Bigl(\frac{p}{2},p(p-1)3^{1-p}\Bigr)$, one obtains
\bea
&&\hspace{-3mm}e^{pV_t^\lambda }|Y_t|^p+\kappa_p \int_t^{\tau_n}e^{pV_s^\lambda}|Y_s|^{p-2}|Z_s|^2ds+\kappa_p \int_t^{\tau_n}\int_E 
e^{pV_s^\lambda}|Y_{s-}|^{p-2}|\psi_s(z)|^2\mu(ds,dz)\nn \\
&&\hspace{-3mm}\leq e^{pV_{\tau_n}^\lambda}|Y_{\tau_n}|^p+\int_t^{\tau_n}e^{pV_s^\lambda}p|Y_s|^{p-2}\Bigl(|Y_s|dN_s^\lambda+\lambda(|Z_s|^2+
||\psi_s||^2_{\mbb{L}^2(E)})ds\Bigr)\nn \\
&&\hspace{-3mm}-\int_t^{\tau_n}e^{pV_s^\lambda}p|Y_s|^{p-2}\langle Y_s,Z_sdW_s\rangle-
\int_t^{\tau_n}\int_E e^{pV_s^\lambda}p|Y_{s-}|^{p-2}\langle Y_{s-},\psi_s(z)\rangle \wt{\mu}(ds,dz).
\label{eq-Yp-Ito}
\eea

Putting $t=0$ and taking expectation give
\bea
&&\mbb{E}\left[\kappa_p\int_0^{\tau_n}e^{pV_s^\lambda}|Y_s|^{p-2}|Z_s|^2ds+\kappa_p \int_0^{\tau_n}\int_E 
e^{pV_s^\lambda}|Y_{s-}|^{p-2}|\psi_s(z)|^2\mu(ds,dz)\right]\nn \\
&&\hspace{-5mm}\leq \mbb{E}\left[e^{pV_{\tau_n}^\lambda}|Y_{\tau_n}|^p+
\int_0^{\tau_n}e^{pV_s^\lambda}p|Y_s|^{p-1}dN_s^\lambda \right]+\lambda \mbb{E}\left[\int_0^{\tau_n}e^{pV_s^\lambda}p|Y_s|^{p-2}\bigl(|Z_s|^2+||\psi_s||^2_{\mbb{L}^2(E)}\bigr)ds\right]~.\nn
\eea
By Lemma~\ref{ap-lemma-nu-vs-mu}, one obtains
\bea
&&\mbb{E}\left[\int_0^{\tau_n} e^{pV_s^\lambda }|Y_s|^{p-2}|Z_s|^2ds+\int_0^{\tau_n}\int_E e^{pV_s^\lambda}|Y_{s-}|^{p-2}|\psi_s(z)|^2\mu(ds,dz)\right]\nn \\
&&\leq C_{p,\lambda}\mbb{E}\left[e^{pV_{\tau_n}^\lambda}|Y_{\tau_n}|^p+\int_0^{\tau_n} e^{pV_s^\lambda }|Y_s|^{p-1}dN_s^\lambda \right]~
\label{eq-Ypm2-Z2}
\eea
by choosing a small $\lambda\in(0,1)$.

Now, applying the Davis inequality (See Chap.I, Sec. 9, Theorem 6 in \cite{Shiryayev}) to (\ref{eq-Yp-Ito}), 
\bea
&&\mbb{E}\Bigl[ ||e^{V^\lambda}Y||^p_{\tau_n}\Bigr]+
\mbb{E}\left[\kappa_p\int_0^{\tau_n}e^{pV_s^\lambda }|Y_s|^{p-2}|Z_s|^2ds+\kappa_p \int_0^{\tau_n}\int_E 
e^{pV_s^\lambda}|Y_{s-}|^{p-2}|\psi_s(z)|^2\mu(ds,dz)\right]\nn \\
&&\leq \mbb{E}\left[e^{pV_{\tau_n}^\lambda}|Y_{\tau_n}|^p+\int_0^{\tau_n}e^{pV_s^\lambda}p|Y_s|^{p-1}dN_s^\lambda \right]
+\lambda \mbb{E}\left[\int_0^{\tau_n}e^{pV_s^\lambda }p|Y_s|^{p-2}\bigl(|Z_s|^2+||\psi_s||^2_{\mbb{L}^2(E)}\bigr)ds\right]\nn \\
&&+C\mbb{E}\Bigl(\int_0^{\tau_n}e^{2pV_s^\lambda}|Y_s|^{2p-2}|Z_s|^2ds\Bigr)^{\frac{1}{2}}
+C\mbb{E}\Bigl(\int_0^{\tau_n}\int_E e^{2pV_s^\lambda}|Y_{s-}|^{2p-2}|\psi_s(z)|^2\mu(ds,dz)\Bigr)^{\frac{1}{2}}~,\nn
\eea
where $C$ is some positive constant.
By Lemma~\ref{ap-lemma-nu-vs-mu}, one can choose $\lambda\in(0,1)$ small enough (depending only on $p$) so that
\bea
&&\mbb{E}\Bigl[ ||e^{V^\lambda}Y||^p_{\tau_n}\Bigr]\leq \mbb{E}\left[e^{pV_{\tau_n}^\lambda}|Y_{\tau_n}|^p
+\int_0^{\tau_n}e^{pV_s^\lambda}p|Y_s|^{p-1}dN_s^\lambda \right]
\nn \\
&&+C\mbb{E}\Bigl(\int_0^{\tau_n}e^{2pV_s^\lambda}|Y_s|^{2p-2}|Z_s|^2ds\Bigr)^{\frac{1}{2}}
+C\mbb{E}\Bigl(\int_0^{\tau_n}\int_E e^{2pV_s^\lambda}|Y_{s-}|^{2p-2}|\psi_s(z)|^2\mu(ds,dz)\Bigr)^{\frac{1}{2}}~.\nn
\eea
By retaking a smaller $\lambda$ in the first step if necessary, one can use a common $\lambda\in(0,1)$
both in the first and second steps.

Note that
\bea
&&C\mbb{E}\Bigl(\int_0^{\tau_n}e^{2pV_s^\lambda}|Y_s|^{2p-2}|Z_s|^2ds\Bigr)^{\frac{1}{2}}\leq C\mbb{E}\left[||e^{V^\lambda} Y||^{\frac{p}{2}}_{\tau_n}\Bigl(\int_0^{\tau_n}e^{pV_s^\lambda}|Y_s|^{p-2}|Z_s|^2ds\Bigr)^{\frac{1}{2}}\right]\nn \\
&&\leq \ep \mbb{E}\Bigl[ ||e^{V^\lambda}Y||^p_{\tau_n}\Bigr]+\frac{C^2}{4\ep}
\mbb{E}\left[\int_0^{\tau_n}e^{pV_s^\lambda}|Y_s|^{p-2}|Z_s|^2ds\right]~,\nn
\eea
and similarly
\bea
&&C\mbb{E}\Bigl(\int_0^{\tau_n}\int_E e^{2pV_s^\lambda}|Y_{s-}|^{2p-2}|\psi_s(z)|^2\mu(ds,dz)\Bigr)^{\frac{1}{2}}\nn \\
&&\leq \ep \mbb{E}\Bigl[||e^{V^\lambda}Y||^p_{\tau_n}\Bigr]+\frac{C^2}{4\ep}
\mbb{E}\left[\int_0^{\tau_n}\int_E e^{pV_s^\lambda}|Y_{s-}|^{p-2}|\psi_s(z)|^2\mu(ds,dz)\right]~.\nn
\eea
Thus, taking $\ep=1/4$, one obtains
\bea
&&\mbb{E}\Bigl[||e^{V^\lambda} Y||^p_{\tau_n}\Bigr]\leq C_p\mbb{E}\left[e^{pV_{\tau_n}^\lambda}|Y_{\tau_n}|^p+\int_0^{\tau_n}e^{pV_s^\lambda}|Y_s|^{p-1}dN_s^\lambda\right]\nn \\
&&+C_p\mbb{E}\left[\int_0^{\tau_n} e^{pV_s^\lambda}|Y_s|^{p-2}|Z_s|^2ds+\int_0^{\tau_n}\int_E e^{pV_s^\lambda}|Y_{s-}|^{p-2}|\psi_s(z)|^2\mu(ds,dz)\right]~.\nn 
\eea
Then the inequality (\ref{eq-Ypm2-Z2}) implies
\bea
\mbb{E}\Bigl[||e^{V^\lambda} Y||^p_{\tau_n}\Bigr]\leq C_{p,\lambda}\mbb{E}\left[e^{pV_{\tau_n}^\lambda}|Y_{\tau_n}|^p+\int_0^{\tau_n}
e^{pV_s^\lambda}|Y_s|^{p-1}dN_s^\lambda\right]~.\nn
\eea
Passing to the limit $\tau_n\rightarrow T$, the monotone convergence in the left and 
the dominated convergence in the right-hand side give
\bea
\mbb{E}\Bigl[||e^{V^\lambda} Y||^p_{T}\Bigr]\leq C_{p,\lambda}\mbb{E}\left[e^{pV_{T}^\lambda}|\wt{\xi}|^p+\int_0^{T}
e^{pV_s^\lambda}|Y_s|^{p-1}dN_s^\lambda
\right]~.\nn
\eea

By Young's inequality,  for an arbitrary $\ep>0$, one has that
\bea
&&\mbb{E}\left[\int_0^T e^{pV_s^\lambda}|Y_s|^{p-1}dN_s^\lambda \right]
\leq \mbb{E}\left[ ||e^{V^\lambda} Y||^{p-1}_T\int_0^T e^{V_s^\lambda}dN_s^\lambda \right]\nn \\
&&\leq \frac{p-1}{p}\ep^{\frac{p}{p-1}}\mbb{E}\Bigl[||e^{V^\lambda} Y||^p_T\Bigr]
+\frac{1}{p\ep^p}\mbb{E}\left[\Bigl(\int_0^T e^{V_s^\lambda}dN_s^\lambda \Bigr)^p\right]~. \nn
\eea
Hence, by taking $\ep$ small, one obtains
\bea
\mbb{E}\Bigl[ ||e^{V^\lambda} Y||^p_T\Bigr]\leq C_{p,\lambda}\mbb{E}\left[e^{pV_T^\lambda}|\wt{\xi}|^p
+\Bigl(\int_0^T e^{V_s^\lambda}dN_s^\lambda \Bigr)^p\right]~, \nn
\eea
Combining with the result (\ref{eq-first-result}) in {\it First step}, one obtains the desired result.
\end{proof}
\end{lemma}

Now, let us introduce the maps $\wt{\xi}^i:\Omega\rightarrow \mbb{R}^m$ and $\wt{f}^i:\Omega\times[0,T]\times \mbb{R}^m\times \mbb{R}^{m\times l}
\times \mbb{L}^2(E,\cale,\nu;\mbb{R}^m)\rightarrow \mbb{R}^m$ with $i\in\{1,2\}$.
\begin{assumption}
\label{assumption-Lipschitz-BSDE}
(i) For $i\in\{1,2\}$, $\wt{\xi}^i$ is $\calf_T$-measurable and the map $(\omega,t)\mapsto \wt{f}^i(\omega,t,\cdot)$ is $\mbb{F}$-progressively measurable.\\
(ii) For every $(y,z,\psi),~(y^\prime,z^\prime,\psi^\prime)
\in \mbb{R}^m\times \mbb{R}^{m\times l}\times \mbb{L}^2(E,\cale,\nu;\mbb{R}^m)$,
there exists a positive  constant $K>0$ such that
\be
|\wt{f}^i(\omega,t,y,z,\psi)-\wt{f}^i(\omega,t,y^\prime,z^\prime,\psi^\prime)|
\leq K\Bigl( |y-y^\prime|+|z-z^\prime|+||\psi-\psi^\prime||_{\mbb{L}^2(E)}\Bigr)\nn 
\ee
$d\mbb{P}\otimes dt$-a.e. in $\Omega\times [0,T]$. \\
(iii) For both $i\in\{1,2\}$, there exists some $p\geq 2$ such that
\bea
\mbb{E}\left[|\wt{\xi}|^p+\Bigl(\int_0^T |\wt{f}(s,0,0,0)|ds\Bigr)^p\right]<\infty~.\nn
\eea
\end{assumption}

\begin{lemma}
\label{ap-lemma-Y-existence}
(a)Under Assumption~\ref{assumption-Lipschitz-BSDE}, the BSDE
\bea
\label{ap-eq-Y}
Y_t^i=\wt{\xi}^i+\int_t^T \wt{f}^i(s,Y_s^i,Z_s^i,\psi_s^i)ds-\int_t^T Z_s^i dW_s-
\int_t^T \int_E \psi_s^i(z)\wt{\mu}(ds,dz)
\eea
has a unique solution $(Y^i,Z^i,\psi^i)$ which belongs to $\mbb{S}^p_m[0,T]\times \mbb{H}^p_{m\times l}[0,T] \times 
\mbb{H}^p_{m,\nu}[0,T]$ satisfying the inequality
\bea
||(Y^i,Z^i,\psi^i)||^p_{\calk^p[0,T]}\leq C_p\mbb{E}\left[|\wt{\xi}|^p+\Bigl(\int_0^T|\wt{f}^i(s,0,0,0)|ds\Bigr)^p\right]
\label{ap-solution-norm}
\eea
where $C_p$ is some positive constant depending only on $(p,K,T)$.
Moreover,
if $A_2^i:=\mbb{E}\Bigl[|\wt{\xi}^i|^2+||\wt{f}^i(\cdot,0)||^2_T\Bigr]<\infty$, then
\be
\label{ap-Yspread-estimate}
\mbb{E}\Bigl[\sup_{s\leq u\leq t}|Y_u^i-Y_s^i|^2\Bigr]\leq C_2
\left[ A_2^i|t-s|^2+\Bigl(\int_s^t |Z_u^i|^2 du\Bigr)+\int_s^t \int_E |\psi_u^i(z)|^2\nu(dz)du\right].
\ee
(b) Fix $\wt{\xi}^1, \wt{\xi}^2\in \mbb{L}^p(\Omega,\calf_T,\mbb{P};\mbb{R}^m)$
and let $(Y^i,Z^i,\psi^i)$ be the solution of (\ref{ap-eq-Y}) for $i\in\{1,2\}$.
Then, for all $t\in[0,T]$, 
\bea
&&\mbb{E}\left[||\del Y||_{[t,T]}^p+\Bigl(\int_t^T |\del Z_s|^2 ds\Bigr)^{p/2}+\Bigl(\int_t^T \int_E |\del\psi_s(z)|^2\mu(ds,dz)\Bigr)^{p/2}
\right]\nn \\
&&+\mbb{E}\left[\Bigl(\int_t^T \int_E |\del\psi_s(z)|^2\nu(dz)ds\Bigr)^{p/2}\right]
  \leq C_p \mbb{E}\left[|\del \xi|^p+
\Bigl(\int_t^T |\del \wt{f}_s| ds\Bigr)^p\right]
\label{ap-delY-estimate}
\eea
where $\del\xi:=\wt{\xi}^1-\wt{\xi}^2$, $\del Y:=Y^1-Y^2$, 
$\del Z:=Z^1-Z^2$, $\del \psi:=\psi^1-\psi^2$ and $\del \wt{f}_\cdot:=
(\wt{f}^1-\wt{f}^2)(\cdot,Y^1_\cdot, Z^1_\cdot, \psi^1_\cdot)$.
\subsubsection*{Remark}
Note that in \cite{Kruse}, the estimates (\ref{ap-solution-norm})
and (\ref{ap-delY-estimate}) are slightly weaker, where the right hand side is given by
$\Bigl(\int_0^T |\wt{f}(s,0,0,0)|^p ds\Bigr)$ instead of $\Bigl(\int_0^T |\wt{f}(s,0,0,0)|ds\Bigr)^p$.
This stems from Lemma~\ref{lemma-monotone-apriori} and can be crucial if one needs to 
apply a fixed-point theorem for a short maturity $T$.
\begin{proof}
Firstly, assume the existence of a solution to (\ref{ap-eq-Y}) such that 
$(Y^i,Z^i,\psi^i)\in\calk^p[0,T]$ for both $i\in\{1,2\}$. One has
\bea
&&\langle Y_s^i,\wt{f}^i(s,Y_s^i,Z_s^i,\psi_s^i)\rangle ds \leq |Y_s^i|\Bigl(|\wt{f}^i(s,0)|+
K\bigl(|Y_s^i|+|Z_s^i|+||\psi_s^i||_{\mbb{L}^2(E)}\bigr)\Bigr)ds \nn \\
&&\quad\leq  |Y_s^i|^2\Bigl(K+\frac{K^2}{2\lambda}\Bigr)ds+|Y_s^i||\wt{f}^i(s,0)|ds+\lambda(|Z_s^i|^2+||\psi^i_s||^2_{\mbb{L}^2(E)})ds\nn
\eea
for $\forall \lambda>0$. One can easily check that Assumption~\ref{assumption-monotone-growth} is satisfied by choosing
\bea
V_t^\lambda:=\Bigl(K+\frac{K^2}{2\lambda}\Bigr)t, \quad N_t^\lambda:=\int_0^t |\wt{f}^i(s,0)|ds, \nn
\eea
for $t\in[0,T]$. Thus Lemma~\ref{lemma-monotone-apriori} proves the inequality (\ref{ap-solution-norm}).

The BDG inequality yields
\bea
\mbb{E}\left[\sup_{u\in[s,t]}|Y_u^i-Y_s^i|^2\right]\leq C_2\mbb{E}\left[
\Bigl(\int_s^t |\wt{f}^i(r,Y_r^i,Z_r^i,\psi_r^i)|dr\Bigr)^2+\int_s^t \Bigl(|Z_r^i|^2+
 ||\psi_r^i||^2_{\mbb{L}^2(E)}\Bigr)dr\right]\nn
\eea
which, together with the estimate (\ref{ap-solution-norm}), proves (\ref{ap-Yspread-estimate}).
For (b), it is easy to check
\bea
|f^1(s,Y_s^1,Z_s^1,\psi_s^1)-f^2(s,Y_s^2,Z_s^2,\psi_s^2)|\leq |\del f_s|+K\Bigl(|\del Y_s|+|\del Z_s|+||\del \psi_s||_{\mbb{L}^2(E)}
\Bigr). \nn
\eea
Thus, Assumption~\ref{assumption-monotone-growth} is satisfied once again for $(\del Y, \del Z, \del \psi)$ by choosing
\bea
V_t^\lambda:=\Bigl(K+\frac{K^2}{2\lambda}\Bigr)t, \quad N_t^\lambda:=\int_0^t |\del f(s)|ds~.\nn
\eea
Therefore, the estimate (\ref{ap-delY-estimate}) immediately follows from Lemma~\ref{lemma-monotone-apriori}.
\\

Now, let us prove the existence in (a). The uniqueness is already proved by (b).
The following is a simple modification of 
Theorem 5.17~\cite{Pardoux-Rascanu} given for a diffusion setup.
Consider a sequence of BSDEs (the superscript $i\in\{1,2\}$ is omitted),  for $n\in\mbb{N}$, 
\bea
Y_t^{n+1}=\wt{\xi}+\int_t^T \wt{f}(s,Y_s^{n},Z_s^{n},\psi_s^n)ds-\int_t^T Z_s^{n+1}dW_s
-\int_t^T \int_E \psi_s^{n+1}(z)\wt{\mu}(ds,dz)~.\nn
\eea
Suppose that $(Y^n,Z^n,\psi^n)\in \calk^p[0,T]$.
Then, from the linear growth property, it is obvious that
\bea
\wt{\xi}+\int_t^T \wt{f}(s,Y_s^{n},Z_s^n,\psi_s^n)ds\in \mbb{L}^p(\Omega,\calf_T,\mbb{P};\mbb{R}^m)~.\nn
\eea
Thus the martingale representation theorem (see, for example, Theorem 5.3.6 in \cite{Applebaum}) implies that there exists a
unique solution $(Y^{n+1},Z^{n+1},\psi^{n+1})\in\calk^p[0,T]$.
Let us define this map as $(Y^{n+1},Z^{n+1},\psi^{n+1})=\Phi(Y^{n},Z^n,\psi^n)$.
Denote $(\del Y^n,\del Z^n,\del \psi^n):=(Y^n-Y^{n-1},Z^n-Z^{n-1},\psi^n-\psi^{n-1})$.
Then (\ref{ap-delY-estimate}) (with a zero Lipschitz constant) implies
\bea
&&||(\del Y^{n+1},\del Z^{n+1},\del \psi^{n+1})||^p_{\calk^p[0,T]} \nn \\
&&\leq C_p \mbb{E}\left[\Bigl(\int_0^T |\wt{f}(s,Y_s^{n},Z_s^n,\psi^n)-\wt{f}(s,Y_s^{n-1},Z_s^{n-1},\psi_s^{n-1})|ds\Bigr)^p
\right] \nn \\
&&\leq C_p^\prime \mbb{E}\left[\Bigl(\int_0^T \bigl[|\del Y^n_s|+|\del Z^n_s|+||\del \psi^n_s||_{\mbb{L}^2(E)}\bigr]ds\Bigr)^p\right]\nn \\
&&\leq C_p^\prime \max(T^p,T^{\frac{p}{2}})||(\del Y^{n},\del Z^{n},\del \psi^{n})||^p_{\calk^p[0,T]}~. 
\eea
Note in particular that $C_p^\prime$ is independent of the terminal condition.
Thus, if the terminal time $T$ is small enough so that $\alpha:=C_p^\prime \max(T^p,T^{\frac{p}{2}})<1$,
then the map $\Phi$ is strictly contracting. 
In this case,  by the fixed point theorem in the Banach space, there exists a 
solution $(Y,Z,\psi)\in \calk^p[0,T]$ to the BSDE (\ref{ap-eq-Y}).
For general $T$, one can consider a time partition $0=T_0<T_1<\cdots<T_N=T$.
By taking $[T_{N-1},T]$ small enough, the above arguments guarantee that there 
exists a solution $(Y,Z,\psi)\in \calk^p[T_{N-1},T]$.  By the uniqueness of the solution,
one can repeat the same procedures for the interval $[T_{N-2},T_{N-1}]$ with 
the new terminal value $Y_{T_{N-1}}$. Repeating $N$ times, one proves the desired result.
\end{proof}
\end{lemma}

The following lemma is useful when one deals with the jumps of finite measure.
\begin{lemma}
\label{ap-lemma-finite}
Suppose $\nu^i(\mbb{R}_0)<\infty$ for every $1\leq i\leq k$. Given $\psi\in\mbb{H}^2_{\nu}[0,T]$, let $M$
be defined by $M_t:=\int_0^t\int_E \psi_s(z)\wt{\mu}(ds,dz)$ on $[0,T]$.
Then, for $\forall p\geq 2$, $k_p ||\psi||^p_{\mbb{H}^p_\nu[0,T]}\leq ||M||^p_{\mbb{S}^p[0,T]}
\leq K_p ||\psi||^p_{\mbb{H}^p_\nu[0,T]}$, where
$k_p, K_p$ are positive constant depend only on $p,\nu(E)$ and $T$.
\begin{proof}
See pp.125 of \cite{Elie}, for example.
\end{proof}
\end{lemma}

\section{Smooth approximation theorem}
\label{sec-smoothing}
In the reminder of the paper, we provide a justification to use smooth coefficients
in the forward-backward SDEs for any numerical approximation purpose.
Since $\ep$ is a perturbation parameter, we can always introduce it so that
all the functions depend smoothly on $\ep$. This is  actually the case for the 
examples used in Sections~\ref{sec-implementation} and \ref{sec-polynomial}.
Thus we concentrate on the other parameters and omit $\ep$ dependence from the functions in the following.
Let us first consider the forward component:
\bea
\label{eq-X-org}
\wt{X}_s=x+\int_t^s \wt{b}(r,\wt{X}_r)dr+\int_t^s \wt{\sigma}(r,\wt{X}_r)dW_r
+\int_t^s \int_E \wt{\gamma}(r,\wt{X}_r,z)\wt{\mu}(dr,dz)~,
\eea
where $x\in\mbb{R}^d$ and $\wt{b}:[0,T]\times \mbb{R}^d\rightarrow \mbb{R}^d$, 
$\wt{\sigma}:[0,T]\times \mbb{R}^d\rightarrow \mbb{R}^{d\times l}$, 
$\wt{\gamma}:[0,T]\times \mbb{R}^d\times E\rightarrow \mbb{R}^{d\times k}$ are measurable functions. We omit the superscripts
denoting the initial data $(t,x)$.

\begin{assumption}
\label{assumption-Xorg}
$\wt{b},\wt{\sigma},\wt{\gamma}$ are continuous in $(t,x,z)$. There exists some positive constant $K$ such that, for every $x,x^\prime \in\mbb{R}^d$, \\
(i) $|\wt{b}(t,x)-\wt{b}(t,x^\prime)|+|\wt{\sigma}(t,x)-\wt{\sigma}(t,x^\prime)|\leq K|x-x^\prime|$ uniformly in $t\in[0,T]$,\\
(ii) $|\wt{\gamma}_j(t,x,z)-\wt{\gamma}_j(t,x^\prime,z)|\leq K\eta(z)|x-x^\prime|$ for $1\leq j\leq k$ uniformly in $(t,z)\in[0,T]\times \mbb{R}_0$, \\
(iii) $||\wt{b}(\cdot,0)||_T+||\wt{\sigma}(\cdot,0)||_T+||\wt{\gamma}(\cdot,0,z)||_T/\eta(z)\leq K$
uniformly in $z\in E$.
\end{assumption}

The regularization technique by the convolution with appropriate mollifiers
gives us the following approximating functions.
\begin{lemma} 
\label{lemma-mollification}
Under Assumption~\ref{assumption-Xorg},
one can choose a sequence of functions $b_n:[0,T]\times \mbb{R}^d\rightarrow \mbb{R}^d$,
$\sigma_n:[0,T]\times \mbb{R}^d\rightarrow \mbb{R}^{d\times l}$, $\gamma_n:[0,T]\times \mbb{R}^d\times E\rightarrow \mbb{R}^{d\times k}$
with $n\in\mbb{N}$, which are continuous in all their arguments, 
infinitely differentiable in $x$ with continuous derivatives, and also satisfy, for each $n\geq 1$ ;\\
(i) for every $m\geq 1$, $|\part_x^m b_n(t,x)|+|\part_x^m \sigma_n(t,x)|+|\part_x^m\gamma_n(t,x,z)|/\eta(z)$ 
is  uniformly bounded in $(t,x,z)\in[0,T]\times \mbb{R}^d\times E$,\\
(ii) for every $(t,x, z)\in [0,T]\times \mbb{R}^d\times E$,  $b_n(t,x), \sigma_n(t,x)$ and $\gamma_n(t,x,z)$
converge pointwise to $\wt{b}(t,x)$, $\wt{\sigma}(t,x)$ and $\wt{\gamma}(t,x,z)$, respectively, \\
(iii) $(b_n, \sigma_n, \gamma_n)$ satisfy the properties in Assumption~\ref{assumption-Xorg}
with some positive constant $K^\prime$ independent of $n$.
\begin{proof}
We consider a sequence of (symmetric) 
mollifiers  $\varrho_n\in \calc_0^{\infty}: \mbb{R}^d\rightarrow \mbb{R}_+$ with compact support satisfying 
$\int_{\mbb{R}^{d}}\varrho_n(x)dx=1$ and $\varrho_n(x)\rightarrow \del(x)$ as $n\rightarrow \infty$ in the space of Schwartz distributions,
where $\del(\cdot)$ is a Dirac delta function.
Let us define intermediate mollified functions as
\be
\bar{b}_n(t,x):=\varrho_n*\wt{b}(t,x), \quad \bar{\sigma}_n(t,x):=\varrho_n*\wt{\sigma}(t,x),
\quad \bar{\gamma}_n(t,x,z):=\varrho_n*\wt{\gamma}(t,x,z) \nn
\ee
where $*$ denotes a convolution with respect to $x$, such as
\be
\bar{b}_n(t,x)=\int_{\mbb{R}^d}\varrho_n(x-y)\wt{b}(t,y)dy=\int_{\mbb{R}^d}\wt{b}(t,x-y)\varrho_n(y)dy~. \nn
\ee
Since $\wt{b},\wt{\sigma},\wt{\gamma}$ are continuous,  every point $x\in \mbb{R}^d$
is a Lebesgue point. Thus, the approximated functions $\bar{b}_n,\bar{\sigma}_n, \bar{\gamma}_n$
are known to converge pointwise to $\wt{b},\wt{\sigma},\wt{\gamma}$
from the Lebesgue differentiation theorem (see, for example, Theorem 8.7 in Igari (1996)~\cite{Igari}
or Theorem~C.19 in Leoni (2009)~\cite{Leoni}).
The Lipschitz property can be shown as, for every $x,x^\prime \in \mbb{R}^d$, 
\bea
&&|\bar{\gamma}_{n,j}(t,x,z)-\bar{\gamma}_{n,j}(t,x^\prime,z)|\leq \int_{\mbb{R}^d}
|\wt{\gamma}_j(t,x-y,z)-\wt{\gamma}_j(t,x^\prime-y,z)|\varrho_n(y)dy\nn \\
&&\qquad \leq K\eta(z)|x-x^\prime|\int_{\mbb{R}^d}\varrho_n(y)dy = K|x-x^\prime|\eta(z) \nn
\eea
and similarly for the others.
It is easy to see that there exists some positive constant $C^\prime$ satisfying
\be
||\bar{b}_n(\cdot,0)||_T+||\bar{\sigma}_n(\cdot,0)||_T+||\bar{\gamma}_n(\cdot,0,z)||_T/\eta(z)\leq C^\prime \nn
\ee
uniformly in $z\in E$ as well as $n\in\mbb{N}$ since $\varrho_n$ has a compact support shrinking to the origin
as $n\rightarrow \infty$.
We prepare another (symmetric) mollifiers $\varsigma_n \in\calc_0^{\infty}$ $:\mbb{R}^d\times E \rightarrow \mbb{R}_+$ 
in the following way:
\bea
\varsigma_n(x,z)=
\begin{cases}
1 & \text{for $|x|+|z| \leq n$} \\
0 & \text{for $|x|+|z| \geq 2n$}
\end{cases} ~.
\label{eq-2nd-mollifier}
\eea
We then define the mollified functions as
\bea
b_n(t,x):=\varsigma_n(x,0) \bar{b}_n(t,x),\quad \sigma_n(t,x):=\varsigma_n(x,0)\bar{\sigma}_n(t,x),\quad
\gamma_n(t,x,z):=\varsigma_n(x,z)\bar{\gamma}_n(t,x,z)~. \nn
\eea
Since they are smooth in $x$ and have compact supports, they have bounded derivatives of all orders with respect to $x$ uniformly in $(t,x, z)$ for each $n$.
The pointwise convergence is clearly preserved.
Lastly, one has to check that there exists a Lipschitz constant $K^\prime$ independent of $n$.
By the construction in (\ref{eq-2nd-mollifier}), one can arrange the mollifier in the following way:
there exists a  positive constant $C$  such that
\be
	\sup_{(x,z)\in\mbb{R}^d\times E}\Bigl|\part_x \varsigma_n(x,z)\Bigr|\leq C/n \nn
\ee
for every $n\in\mbb{N}$. Then, for $\forall n\in\mbb{N}$, one sees
\bea
|\part_x \gamma_n(t,x,z)|&\leq& \varsigma_n(x,z)|\part_x \bar{\gamma}_n(t,x,z)|+|\part_x \varsigma_n(x,z)||\bar{\gamma}_n(t,x,z)|\nn \\
&\leq &K\eta(z)+\eta(z)C/n(C^\prime+K(2n))\leq K^\prime \eta(z)\nn
\eea
uniformly in $(t,x,z)$. Here, we have used the fact that $\part_x\varsigma_n(x,z)$ vanishes when $|x|\geq 2n$ and the linear growth property of $\bar{\gamma}_n$. 
One can similarly check $|\part_x b_n(t,x)|, |\part_x \sigma_n(t,x)|\leq K^\prime$ for $\forall n\in \mbb{N}$.
The property (iii) of Assumption \ref{assumption-Xorg} is obviously preserved in the second mollification.
\end{proof}
\end{lemma}

This yields the following result.
\begin{theorem}
\label{theorem-X-mollification}
Under Assumption~\ref{assumption-Xorg}, consider the process $\wt{X}$ of (\ref{eq-X-org}) 
and the sequence of processes $(X^n_s,s\in[t,T])_{n\geq 1}$ defined by
\bea
\label{eq-X-mollified}
X^n_s=x+\int_t^s b_n(r,X_r^n)dr+\int_t^s \sigma_n(r,X^n_r)dW_r+\int_t^s \int_E \gamma_n(r,X^n_r,z)\wt{\mu}(dr,dz) 
\eea
with $b_n,\sigma_n$ and $\gamma_n$ given in Lemma~\ref{lemma-mollification}.  Then, there exist
unique solutions $\wt{X}, X^{n}$ in $\mbb{S}^p[t,T]~\forall p\geq 2$.
Moreover, the following relation holds
\be
 \lim_{n\rightarrow \infty}\mbb{E}\Bigl[||\wt{X}-X^n||_{[t,T]}^p\Bigr]=0 \nn
\ee
for $\forall p\geq 2$.
\begin{proof}
The existence of the unique solution for (\ref{eq-X-org}) as well as (\ref{eq-X-mollified}) in $\mbb{S}^p$ for $\forall p\geq 2$ 
is clear from Lemma~\ref{ap-lemma-X-existence}. We also have, for $\forall p\geq 2$,
\bea
||\wt{X}-X^n||^p_{\mbb{S}^p}&\leq& C_p \mbb{E}\left[ \Bigl(\int_t^T |\del \wt{b}_n(r,\wt{X}_r)|dr\Bigr)^p
+\Bigl(\int_t^T |\del \wt{\sigma}_n(r,\wt{X}_r)|^2dr\Bigr)^{p/2}+\int_t^T |\del L^n_r|^p dr\right] \nn
\eea
where $\del \wt{b}_n:=\wt{b}-b_n$, $\del \wt{\sigma}_n:=\wt{\sigma}-\sigma_n$.
Furthermore $\del L^n$ is a predictable process satisfying $|\del \wt{\gamma}_n|(t,\wt{X}_{t-},z)\leq \del L^n_t \eta(z)$,
$d\mbb{P}\otimes dt$-a.e. in $\Omega\times [0,T]$, where $\del \wt{\gamma}_n:=\wt{\gamma}-\gamma_n$.
We can take $\del L^n$ such that $\int_t^T \mbb{E}|\del L^n_r|^p dr<\infty$,
since we have $|\del \wt{\gamma}_n|(s,\wt{X}_{s-},z) \leq 2K(1+|\wt{X}_{s-}|)\eta(z)$ in the current setup.
See also the related discussion in Lemma~\ref{ap-lemma-X-existence}.

Note that $C_p$ is independent of $n$ thanks to  Lemma~\ref{lemma-mollification} (iii).
Due to the linear growth property, the inside of the expectation is dominated by $C(1+||\wt{X}||_{[t,T]}^p)$ with some positive constant $C$ independent of $n$.
From Lemma~\ref{lemma-mollification} (ii), $(\del \wt{b}_n,\del\wt{\sigma}_n,\del \wt{\gamma}_n)$ converge pointwise to zero.
Thus, one can also take a sequence of $(\del L^n, n\in\mbb{N})$ converging pointwise to zero.
Since $\wt{X}\in\mbb{S}^p$ for $\forall p\geq 2$, 
the dominated convergence theorem give the desired result in the limit 
$n\rightarrow \infty$.~\footnote{In $p=2$, one can see more directly $||\wt{X}-X^n||^2_{\mbb{S}^2}\rightarrow 0$
since the integral of $\del L^n$ can be replaced by that of $\del{\wt{\gamma}}_n$ (See a remark 
below Lemma~\ref{ap-lemma-X-existence}.). Taking an appropriate subsequence if necessary,
one can also show that $(X^n_s,s\in[t,T])_{n\geq 1}$ is
almost surely uniformly convergent to $(\wt{X}_s,s\in[t,T])$
by the Borel-Cantelli lemma.}
\end{proof}
\end{theorem}

The above result implies that by choosing a large enough $n$
one can work on $X^n$ that is an arbitrary accurate approximation in the $\mbb{S}^p$ sense  of the original process $\wt{X}$, 
and involves only smooth coefficients $(b_n,\sigma_n,\gamma_n)$. This conclusion can be extended to the forward-backward system.
Consider the BSDE driven by $\wt{X}$;
\bea
\label{eq-Y-org}
&&\wt{Y}_s=\wt{\xi}(\wt{X}_T)+\int_s^T\wt{f}\Bigl(r,\wt{X}_r,\wt{Y}_r,\wt{Z}_r,\int_{\mbb{R}_0}\rho(z)\wt{\psi}_r(z)\nu(dz)\Bigr)dr\nn \\
&&\qquad -\int_s^T \wt{Z}_rdW_r-\int_s^T \int_E \wt{\psi}_r(z)\wt{\mu}(dr,dz)
\eea
for $s\in[t,T]$ where $\wt{\xi}:\mbb{R}^d\rightarrow \mbb{R}^m$,
$\wt{f}:[0,T]\times \mbb{R}^d\times \mbb{R}^m\times \mbb{R}^{m\times l}\times \mbb{R}^{m\times k}\rightarrow \mbb{R}^m$ are 
measurable functions and
$\rho$ is defined as before. 
\begin{assumption}
\label{assumption-Y-mollification}
The functions $\wt{\xi}$ and $\wt{f}$ are continuous in all their arguments. There
exist some positive constants $K,q\geq 0$ such that \\
(i)$|\wt{\xi}(x)|+|\wt{f}(t,x,0,0,0)|\leq K(1+|x|^q)$ for every $x\in\mbb{R}^d$ uniformly in $t \in[0,T]$. \\
(ii) $|\wt{f}(t,x,y,z,u)-\wt{f}(t,x,y^\prime,z^\prime,u^\prime)|\leq K(|y-y^\prime|+|z-z^\prime|+|u-u^\prime|)$
for every $(y,z,u), (y^\prime,z^\prime,u^\prime)\in \mbb{R}^m\times \mbb{R}^{m\times l} \times \mbb{R}^{m \times k}$
uniformly in $(t,x)\in[0,T]\times \mbb{R}^d$.
\end{assumption}

\begin{lemma}
\label{lemma-Y-mollification}
Under Assumption~\ref{assumption-Y-mollification}, one can choose a sequence of functions $\xi_n:\mbb{R}^d\rightarrow \mbb{R}^m$,
$f_n: [0,T]\times \mbb{R}^d\times \mbb{R}^m \times \mbb{R}^{m\times l}\times \mbb{R}^{m\times k}\rightarrow \mbb{R}^m$
with $n\in\mbb{N}$, which are continuous in all their arguments, 
infinitely differentiable in $(x,y,z,u)$ with continuous derivatives, and also satisfy, for each $n\geq 1$;
 \\
(i) for every $i\geq 1$, all the $i$th order partial derivatives of $(\xi_n, f_n)$ are uniformly bounded 
in $(t,x,y,z,u)\in[0,T]\times \mbb{R}^d\times \mbb{R}^m\times \mbb{R}^{m\times l} \times \mbb{R}^{m\times k}$, \\
(ii) for every $(t,x,y,z,u)\in[0,T]\times \mbb{R}^d\times \mbb{R}^m\times  \mbb{R}^{m\times l}\times \mbb{R}^{m\times k}$,
$\xi_n$ and $f_n$ converge pointwise to $\wt{\xi}$ and $\wt{f}$, respectively, \\
(iii) $(\xi_n, f_n)$ satisfy  Assumption~\ref{assumption-Y-mollification} with some positive constant $K^{\prime\prime}$ 
independent of $n$.
\begin{proof}
The first step of the mollification can be done exactly the same way as in  Lemma~\ref{lemma-mollification},
which gives us $\bar{\xi}_n(x)$ and $\bar{f}_n(t,x,\hat{\Theta}):=\bar{f}_n(t,x,y,z,u)$.
In order to achieve the property (iii), one has to take care of the 
polynomial growth property of the driver with respect to $x$.
One can take the second sequence of mollifiers as 
\bea
\varsigma_n(x,\hat{\Theta})=\begin{cases}
1 & \text{for $|x|^q+|\hat{\Theta}| \leq n$} \\
0 & \text{for $|x|^q+|\hat{\Theta}| \geq 2n$}
\end{cases} \nn
\eea
and then control their first derivatives, with some positive constant $C$, by
\be
\sup_{(x,\hat{\Theta})\in\mbb{R}^d\times \mbb{R}^{m(1+l+k)}}|\part_{\hat{\Theta}}\varsigma_n(x,\hat{\Theta})|\leq C/n \nn
\ee
for $\forall n\in\mbb{N}$. Then, one can check that 
\be
\xi_n(x):=\varsigma_n(x,0)\bar{\xi}_n(x),\quad f_n(t,x,\hat{\Theta}):=\varsigma_n(x,\hat{\Theta})\bar{f}_n(t,x,\hat{\Theta}) \nn
\ee
satisfy the desired property similarly as in Lemma~\ref{lemma-mollification}.
\end{proof}
\end{lemma}

Finally, we obtain the main approximation theorem.
\begin{theorem}
\label{theorem-Y-mollification}
Under Assumptions~\ref{assumption-Xorg} and 
\ref{assumption-Y-mollification}, consider the process $(\wt{Y},\wt{Z},\wt{\psi})$ of (\ref{eq-Y-org}) and
the sequence of processes $(Y^{n,m}_s,Z_s^{n,m},\psi_s^{n,m})_{s\in[t,T]}, (n,m)\in\{1,2,\cdots\}$ 
defined as the solution to following BSDE
\bea
&&Y_s^{n,m}=\xi_m(X^n_T)+\int_s^T f_m\Bigl(r,X_r^n,Y_r^{n,m},Z_r^{n,m},\int_{\mbb{R}_0}\rho(z) \psi_r^{n,m}(z)\nu(dz)
\Bigr)dr\nn \\
&&\qquad -\int_s^T Z_r^{n,m}dW_r -\int_s^T \int_E \psi_r^{n,m}(z)\wt{\mu}(dr,dz)
\eea
where $X^n$ is the solution of (\ref{eq-X-mollified}), $(\xi_m,f_m)$ are the mollified functions given in 
Lemma~\ref{lemma-Y-mollification}.
Then, there exist unique solutions $(\wt{Y},\wt{Z},\wt{\psi})$, $(Y^{n,m},Z^{n,m},\psi^{n,m})_{n,m\geq 1}\in \calk^p[t,T]~\forall p\geq 2$. Moreover, the following relation holds
\be
\lim_{m\rightarrow \infty} \lim_{n\rightarrow \infty} \Bigl|\Bigl|(\del Y^{n,m},\del Z^{n,m},
\del \psi^{n,m})\Bigr|\Bigr|_{\calk^p[t,T]}=0~~ \forall p\geq 2\nn
\ee
 where $\del Y^{n,m}:=\wt{Y}-Y^{n,m}$, $\del Z^{n,m}:=\wt{Z}-Z^{n,m}$ and $\del \psi^{n,m}:=\wt{\psi}-\psi^{n,m}$.
\begin{proof}
The existence of the unique solution $(\wt{Y},\wt{Z},\wt{\psi})$ and 
$(Y^{n,m},Z^{n,m},\psi^{n,m})$ in $\calk^p$ for $\forall p\geq 2$ is clear from Lemma~\ref{ap-lemma-Y-existence}.
We have, for  $\forall p\geq 2$, 
\bea
\Bigl|\Bigl|(\del Y^{n,m},\del Z^{n,m},\del \psi^{n,m})\Bigr|\Bigr|^p_{\calk^p[t,T]} \leq C_p \mbb{E}\left[|\del \xi^{n,m}|^p+\Bigl(\int_t^T |\del f^{n,m}(r)|dr\Bigr)^p\right] \nn
\eea
by the stability result, where $\del \xi^{n,m}:=\wt{\xi}(\wt{X}_T)-\xi_m(X_T^n)$ and
\bea
\del f^{n,m}(r)&:=&\wt{f}\Bigl(r,\wt{X}_r,\wt{Y}_r,\wt{Z}_r,\int_{\mbb{R}_0}\rho(z)\wt{\psi}_r(z)\nu(dz)\Bigr)\nn \\
&& -f_m\Bigl(r,X_r^n,\wt{Y}_r,\wt{Z}_r,\int_{\mbb{R}_0}\rho(z)\wt{\psi}_r(z)\nu(dz)\Bigr)~. \nn
\eea
Firstly, let us fix $m$. Since $\part_x \xi_m$ and $\part_x f_m $ are bounded,
the result of Theorem~\ref{theorem-X-mollification} yields
\bea
\lim_{n\rightarrow \infty}\Bigl|\Bigl|(\del Y^{n,m},\del Z^{n,m},\del \psi^{n,m})\Bigr|\Bigr|^p_{\calk^p[t,T]}  \leq C_p \mbb{E}\left[ |\del \xi^m|^p+
\Bigl(\int_t^T |\del f^m (r,\wt{\Theta}_r)|dr\Bigr)^p\right] \nn
\eea
with $\del \xi^m:=\wt{\xi}(\wt{X}_T)-\xi_m(\wt{X}_T)$ and $
\del f^m(r,\wt{\Theta}_r):=(\wt{f}-f_m)\Bigl(r,\wt{X}_r,\wt{Y}_r,\wt{Z}_r,\int_{\mbb{R}_0}\rho(z)\wt{\psi}_r(z)\nu(dz)\Bigr)$.
Since $\wt{\Theta}\in \mbb{S}^p\times \calk^p$ for $\forall p\geq 2$ and
$(\wt{f},f_m)$ have the linear growth in $(y,z,u)$ and the polynomial growth in $x$
with proportional coefficients independent of $m$, 
passing to the limit $m\rightarrow \infty$ yields 
the desired result from the pointwise convergence of the mollified functions and 
the dominated convergence theorem.
Notice also that one can  achieve the 
same convergence with the flipped order of limits $\lim_{n\rightarrow \infty}\lim_{m\rightarrow \infty}$
by using the fact that  $(X^n_s,s\in[t,T])_{n\in\mbb{N}}$ is almost surely uniformly convergent to $(\wt{X}_s,s\in[t,T])$ 
by taking an appropriate subsequence if necessary.
\end{proof}
\end{theorem}

Theorems~\ref{theorem-X-mollification} and \ref{theorem-Y-mollification}
imply that one can work on the process $\Theta^n$ defined by the smooth coefficients $(b_n,\sigma_n, \gamma_n, \xi_n, f_n)$
as an arbitrary accurate approximation in the $\mbb{S}^p\times \calk^p$ sense of the original one $\wt{\Theta}$, 
which only satisfies Assumptions~\ref{assumption-Xorg} and \ref{assumption-Y-mollification}.
In fact, we can weaken the assumptions further.
There is no difficulty to add discontinuities to $\wt{\xi}$ and $\wt{f}$ with respect to $x$
as long as they are all Lebesgue points.
If we only assume, in addition to the polynomial growth condition,  that $(\wt{\xi}, \wt{f})$ is Borel measurable,
then $(\xi_m,f_m)$ converges to $(\wt{\xi},\wt{f})$ only $dx$-a.e. (and hence $(\wt{\xi},\wt{f})$ does not have
Lebesgue points everywhere) in general.
As long as the forward process $\wt{X}$ has no mass on this null set in $dx$, the same conclusion will hold.

\section*{Acknowledgement}
The research is partially supported by Center for Advanced Research in Finance (CARF) and
JSPS KAKENHI (Grant Number 25380389).




\begin{thebibliography}{99}
\small 
\bibitem{Applebaum}
Applebaum, D., 2009, {\it L\'evy Processes and Stochastic Calculus (2nd edition)},
Cambridge studies in advanced mathematics, Cambridge University Press, LN.

\bibitem{Barles}
Barles, G., Buckdahn, R. and Pardoux, E., 1997, {\it Backward stochastic differential equations
and integral-partial differential equations},
Stochastics and Stochastic Reports, Vol. 60, 57-83.


\bibitem{Bender-Denk}
Bender, C. and Denk, R., 2007, {\it A forward scheme for backward SDE},
Stochastic Processes and their Applications, 117, 12, 1793-1832.


\bibitem{Bichteler}
Bichteler, K., Gravereaux, J. and Jacod, J., 1987, {\it Malliavin calculus for processes with jumps},
Stochastics Monographs,  Gordon and Breach science publishers, LN.

\bibitem{Bismut}
Bismut, J.M., 1973, {\it Conjugate convex functions in optimal stochastic control}, 
J. Math. Anal. Apl. 44, 384-404.



\bibitem{Bouchard-Elie}
Bouchard, B. and Elie, R., 2008, {\it Discrete-time approximation of decoupled Forward-Backward SDE with Jumps},
Stochastic Processes and their Applications, 118, 53-75.

\bibitem{Bouchard-Touzi}
Bouchard, B. and Touzi, N., 2004, {\it Discrete-time approximation and Monte-Carlo
simulation of backward stochastic differential equations},
Stochastic Processes and their Applications, 111, 2, 175-206.


\bibitem{Crepey}
Cr\'epey, S. and Bielecki, T.R. with an introductory dialogue by Damiano Brigo, 2014,
{\it Counterparty Risk and Funding}, CRC press, Taylor \& Francis Group, NY.

\bibitem{Crepey-Matoussi}
Crepey, S. and Matoussi, A., 2008, {\it Reflected and doubly reflected BSDEs with jumps:
A priori estimates and comparison}, The Annals of Applied Probability, 18, 5, 2041-2069.

\bibitem{Crepey-Song}
Cr\'epey, S. and Song, S., 2015, {\it Counterparty Risk and Funding: Immersion and Beyond},
Working paper.

\bibitem{Cvitanic}
Cvitani\'c, J. and Zhang, J., 2013, {\it Contract theory in continuous-time methods},
Springer,Berlin.


\bibitem{Delong}
Delong, L., 2013, {\it Backward Stochastic Differential Equations with Jumps and Their
Actuarial and Financial Applications}, Springer-Verlag, LN.

\bibitem{Delong-Imkeller}
Delong, L. and Imkeller, P., 2010, {\it On Malliavin's differentiability of BSDEs with 
time delayed generators driven by Brownian motions and Poisson random measures},
Stochastic Processes and their Applications, 120, 9, 1748-1775.

\bibitem{Nunno}
Di Nunno, G., Oksendal, B. and Proske, F., 2009, {\it Malliavin Calculus for L\'evy Processes 
with Applications to Finance}, Springer, NY.

\bibitem{Dzhaparidze}
Dzhaparidze, K. and Valkeila, E., 1990, {\it On the Hellinger type distances for filtered
experiments}, Probab. Theory Related Fields, 85, 1, 105-117.

\bibitem{Elie}
Elie, R., 2006, {\it Contr\^ole stochastique et m\'ethodes num\'eriques en finance math\'ematique},
Ph.D Thesis, University Paris-Dauphine.

\bibitem{ElKaroui-Mazliak}
El Karoui, N. and Mazliak, L. (eds.), 1997, {\it Backward stochastic differential equations},
Addison Wesley Longman Limited, U.S..

\bibitem{ElKaroui}
El Karoui, N., Peng, S. and Quenez, M.-C., 1997, {\it Backward stochastic differential equations in finance},
Mathematical Finance, 7(1), 1-71.

\bibitem{F-Filtering}
Fujii, M., 2014, {\it Momentum-space approach to asymptotic expansion for stochastic filtering},
Annals of the Institute of Statistical Mathematics, Vol. 66, 93-120.

\bibitem{Fujii-poly}
Fujii, M., 2015, {\it A polynomial scheme of asymptotic expansion for backward SDEs and option pricing},
Quantitative Finance, in press.

\bibitem{Fujii-mm}
Fujii, M., 2015, {\it Optimal position management for a market-maker with stochastic price impacts},
Working paper, CARF-F-360. 

\bibitem{FT-analytic}
Fujii, M. and Takahashi, A., 2012, {\it Analytical approximation for non-linear FBSDEs with perturbation 
scheme}, International Journal of Theoretical and Applied Finance, 15, 5, 1250034 (24).



\bibitem{FT-particle}
Fujii, M. and Takahashi, A., 2015b, {\it Perturbative Expansion Technique for Non-linear FBSDEs 
with Interacting Particle Method}, Asia-Pacific Financial Markets, 22, 3, 283-304.


\bibitem{FT-short-term}
Fujii, M. and Takahashi, A., 2015, {\it Solving Backward Stochastic Differential Equations by Connecting the Short-term Expansions},
Working paper, CARF-F-387.


\bibitem{FT-qg-jumps}
Fujii, M. and Takahashi, A., 2017, {\it Quadratic-exponential growth BSDEs with Jumps and their Malliavin's Differentiability},
Stochastic processes and their applications, in press.
https://doi.org/10.1016/j.spa.2017.09.002.

\bibitem{Gikhman}
Gikhman, I.I. and Skorohod, A.V., 1972, {\it Stochastic differential equations},
Springer, Berlin.


\bibitem{Gobet}
Gobet, E., Lemor, J.P. and Warin, X., 2005, {\it A regression-based Monte Carlo method to 
solve backward stochastic differential equations},
The Annals of Applied Probability, 15, 3, 2172-2202.

\bibitem{Jeanblanc}
Hamad\`ene, S. and Jeanblanc, M., 2007, {\it On the stating and stopping problem: Application
in reversible investments},
Mathematics of Operations Research, Vol. 32, No. 1, pp. 182-192.

\bibitem{Henry}
Henry-Labord\`ere, P., 2012, {\it Cutting CVA's complexity},
Risk Magazine, July Issue, 67-73.

\bibitem{He-Wang-Yan}
He, S., Wang, J. and Yan, J., {\it Semimartingale Theory and Stochastic Calculus},
CRC Press, London, UK.

\bibitem{Igari}
Igari, S., 1996, {\it Introduction of Real Analysis},
Iwanami publisher, in Japanese.


\bibitem{Kazamaki}
Kazamaki, N., 1979, {\it A sufficient condition for the uniform integrability of
exponential martingales}, Math. Rep. Toyama Univ. 2, 1-11. MR-0542374.


\bibitem{Kruse}
Kruse, T. and Popier, T., 2015, {\it BSDEs with monotone generator driven by Brownian and
Poisson noises in a general filtration}, 
Stochastics, 88(4):491-539, 2016

\bibitem{Lejay}
Lejay, A., Mordecki, E. and Torres, S., 2014, {\it Numerical approximation of backward stochastic differential equations
with jumps}, HAL: inria-00357992.

\bibitem{Leoni}
Leoni, G., 2009, {\it A First Course in Sobolev Spaces},
Graduate Studies in Mathematics, Vol. 105, American Mathematical Society, US.

\bibitem{Lim}
Lim, A.E., {\it Quadratic Hedging and Mean-Variance portfolio selection
with random parameters in an incomplete market},
Mathematics of Operations Research, Vol. 29, No. 1, 132-161.

\bibitem{Shiryayev}
Liptser, R.Sh. and Shiryayev, A.N., 1989, {\it Theory of Martingales},
Mathematics and Its Applications, Kluwer Academic Publishers, LN.


\bibitem{Ma-Yong}
Ma, J. and Yong, J., 2000, {\it Forward-backward stochastic differential equations and 
their applications},Springer, Berlin.


\bibitem{Ma-Zhang}
Ma, J. and Zhang, J., 2002, {\it Representation theorems for backward stochastic differential 
equations}, The annals of applied probability, 12, 4, 1390-1418.

\bibitem{Mania}
Mania, M. and Tevzadze, R., 2003, {\it Backward stochastic PDE and imperfect hedging},
International Journal of Theoretical and Applied Finance, Vol. 6, 7, 663-692.

\bibitem{Morlais}
Morlais, M-A., 2010, {\it A new existence result for quadratic BSDEs with jumps
with application to the utility maximization problem},
Stochastic processes and their applications, 120, 1966-1995.

\bibitem{Pardoux-Peng}
Pardoux, E. and Peng, S., 1990, {\it Adapted solution of a backward stochastic differential 
equations}, Systems Control Lett., 14, 55-61.

\bibitem{Pardoux-Peng-lec}
Pardoux, E. and Peng, S., 1992, {\it Backward stochastic differential equations
and quasilinear parabolic partial differential equations}, 
Lecture Notes in Control and Inform. Sci. 176, 200-217,
Springer, New York.

\bibitem{Pardoux-Rascanu}
Pardoux, E. and Rascanu, A., 2014, {\it Stochastic Differential Equations, Backward SDEs,
Partial Differential Equations}, Springer International Publishing, Switzerland.

\bibitem{Petrou}
Petrou, E., 2008, {\it Malliavin calculus in L\'evy spaces and applications to finance},
Electric Journal of Probability, 13, 27, 852-879.

\bibitem{Pham}
Pham, H., 2010, {\it Stochastic control under progressive enlargement of filtrations
and applications to multiple defaults risk management},
Stochastic processes and their Applications, 120, 1795-1820.


\bibitem{Protter}
Protter, P.E., 2005, {\it Stochastic Integration and Differential Equations},
Springer, NY.

\bibitem{Quenez-S}
Quenez, M.C. and Sulem, A., 2013, {\it BSDEs with jumps, optimization and applications
to dynamic risk measures}, Stochastic Processes and their Applications, 123, 8, 3328-3357.

\bibitem{Royer}
Royer, M., 2006, {\it Backward stochastic differential equations with jumps and related
non-linear expectations},
Stochastic process and their applications, 116, 1358-1376.


\bibitem{T-review}
Takahashi, A., 2015, {\it Asymptotic Expansion Approach in Finance}, in 
P Fritz, J Gatheral, A Gulisashvili, A Jacquier and J Teichmann (eds.),
{\it Large Deviations and Asymptotic Methods in Finance}, pp. 345-411, Springer.

\bibitem{TY-MOR}
Takahashi, A. and Yamada, T., 2015, {\it On error estimates for asymptotic expansions
with Malliavin weights: Application to stochastic volatility model},
Mathematics of Operations Research, Vol. 40, No. 3, pp. 513-541.

\bibitem{TY-AAP}
Takahashi, A. and Yamada, T., 2015, {\it A weak approximation with asymptotic expansion
and multidimensional Malliavin weights},
Annals of Applied Probability, in press.

\bibitem{Takahashi-Yamada}
Takahashi, A. and Yamada, T., 2015, {\it An asymptotic expansion of forward-backward SDEs with a 
perturbed driver}, Forthcoming in International Journal of Financial Engineering.




\bibitem{Touzi}
Touzi, N., 2013, {\it Optimal stochastic control, stochastic target problems, 
and backward SDEs}, Springer, New York.

\bibitem{ZhangJ}
Zhang, J., 2004, {\it A numerical scheme for bsdes},
The annals of applied probability, 14, 1, 459-488.
\end{thebibliography}
\end{document}